%% file: main.tex
\documentclass[journal]{IEEEtran}

\usepackage[utf8]{inputenc}
\usepackage{amsmath,amssymb,amsfonts,amsthm,mathrsfs}
\usepackage[british]{babel}

\usepackage{algorithm,algorithmic}

\usepackage{graphicx}
\usepackage{textcomp}

\usepackage{subcaption}

\usepackage{enumerate}
\usepackage{float}
\usepackage[acronym]{glossaries}
\usepackage{enumitem}
\usepackage{multirow}
\usepackage{times}
\usepackage{mathtools}
\usepackage[compress]{cite}
\usepackage{etoolbox}
\usepackage{xcolor}

\input{acronyms}

\newcommand{\bs}{\boldsymbol}
\DeclareMathOperator{\SINR}{SINR}
\DeclareMathOperator{\diag}{diag}

\def\BibTeX{{\rm B\kern-.05em{\sc i\kern-.025em b}\kern-.08em
    T\kern-.1667em\lower.7ex\hbox{E}\kern-.125emX}}

\begin{document}

\title{Energy-Efficient Access-Point Sleep-Mode Techniques for Cell-Free mmWave Massive MIMO Networks with Non-uniform Spatial Traffic Density}

\author{
Jan~Garc\'ia-Morales,~\IEEEmembership{Member, IEEE},
Guillem~Femenias,~\IEEEmembership{Senior Member, IEEE},
and
Felip~Riera-Palou,~\IEEEmembership{Senior Member, IEEE}
\thanks{
Jan García-Morales, Guillem Femenias and Felip Riera-Palou are with the Mobile Communications Group, University of the Balearic Islands, Palma 07122, Illes Balears, Spain (e-mail: \{jan.garcia,guillem.femenias,felip.riera\}@uib.es).
}
\thanks{
This work was funded in part by the Agencia Estatal de Investigaci\'on and Fondo Europeo de Desarrollo Regional (AEI/FEDER, UE) and Ministerio de Econom\'ia y Competitividad (MINECO), Spain, under project TERESA (TEC2017-90093-C3-3-R).
}
\thanks{
Author Accepted Manuscript of the article: J. García-Morales, G. Femenias, and F. Riera-Palou,
``Energy-Efficient Access-Point Sleep-Mode Techniques for Cell-Free mmWave Massive MIMO Networks with Non-Uniform Spatial Traffic Density,''
in \emph{IEEE Access}, vol.~8, pp.~137587--137605, 2020.
The final published version is available at:

https://doi.org/10.1109/ACCESS.2020.3012199
}
}

\maketitle

\begin{abstract}
Cell-free massive multiple-input multiple-output (MIMO) is a novel beyond 5G (B5G) and 6G paradigm that, through the use of a common central processing unit (CPU), coordinates a large number of distributed access points (APs) to coherently serve mobile stations (MSs) on the same time/frequency resource. By exploiting the characteristics of new less-congested millimeter wave (mmWave) frequency bands, these networks can improve the overall system spectral and energy efficiencies by using low-complexity hybrid precoders/decoders. For this purpose, the system must be correctly dimensioned to provide the required quality of service (QoS) to MSs under different traffic load conditions. However, only heavy traffic load conditions are usually taken into account when analysing these networks and, thus, many APs might be underutilized during low traffic load periods, leading to an inefficient use of resources and waste of energy. Aiming at the implementation of energy-efficient AP switch on/off strategies, several approaches have been proposed in the literature that only consider rather unrealistic uniform spatial traffic distribution in the whole coverage area. Unlike prior works, this paper proposes energy efficient AP sleep-mode techniques for cell-free mmWave massive MIMO networks that are able to capture the inhomogeneous nature of spatial traffic distribution in realistic wireless networks. The proposed framework considers, analyzes and compares different AP switch ON-OFF (ASO) strategies that, based on the use of goodness-of-fit (GoF) tests, are specifically designed to dynamically turn on/off APs to adapt to both the number and the statistical distribution of MSs in the network. Numerical results show that the use of properly designed GoF-based ASO strategies under a non-uniform spatial traffic distribution can serve to considerably improve the achievable energy efficiency.
\end{abstract}


\section{Introduction}

\subsection{Motivation and previous work}

\IEEEPARstart{C}{ell-free} massive \gls{MIMO}, originally proposed by Ngo \textit{et al.} in \cite{Ngo15}, is a novel wireless networking paradigm currently being investigated in the context of \gls{B5G} and \gls{6G} mobile communications. One of the defining and foremost important features of cell-free massive \gls{MIMO} is the replacement of the classical cell-based structure, pervasive in current wireless networks, by a large number of randomly deployed \glspl{AP} scattered throughout the coverage area, all connected to a \gls{CPU} that coordinates the communications \cite{Ngo18}. Note that this architecture eliminates the concept of cell altogether, hence also avoiding cell-edge performance issues. Importantly, the large number of \glspl{AP}, each potentially equipped with multiple antennas, makes cell-free massive \gls{MIMO} a distributed form of the \textit{classical} centralized massive \gls{MIMO}, thus inheriting many of its attractive features such as the channel hardening and favourable propagation effects through the implementation of simple signal processing at both transmission ends. Recent research has shown that bringing the \gls{RF} front-end closer to the users while allowing certain operations to be conducted centrally (i.e., power and pilot allocation) significantly outperforms conventional architectures such as small cells or centralized \gls{MIMO} strategies in providing a uniform \gls{QoS} throughout the network \cite{Ngo17,Bjornson19}. Moreover, the cell-free massive \gls{MIMO} paradigm allows for a variety of trade-offs in terms of performance, complexity and fronthaul link requirements to be implemented depending on the capabilities of both the \gls{CPU} and the \glspl{AP} and/or the capacities of the fronthaul links connecting the \glspl{AP} to the \gls{CPU}. In particular, it was shown in \cite{Nayebi17,Bjornson19} that conducting the precoding operation at the \gls{CPU} in a cooperative centralized manner while relying on instantaneous \gls{CSI} leads to a considerable improvement over simpler \gls{AP}-based non-cooperative beamforming strategies.

On another front, the rapid increase in mobile data demand over the last decade has virtually filled up most of the conventional mobile/wireless frequency bands (from 300 MHz to 6 GHz), leading to the so-called spectrum crunch and the need to explore alternative frequency regions to accommodate future standards/services \cite{Rappaport13,Boccardi14,Akdeniz14}. The so-called \gls{mmWave} band, located between 6 and 300 GHz, has emerged as the most viable candidate in the short term and in fact it is already playing a significant role in the roll-out of \gls{5G} networks \cite{Rappaport17}, a trend likely to be fully developed in the context of B5G and 6G systems. Despite the large chunks of available spectrum at \gls{mmWave} bands, this portion of spectrum presents important challenges from a communication point of view, most notably, a very severe propagation pathloss. Large antenna arrays with many radiating elements can be used to effectively implement \gls{mmWave} massive \gls{MIMO} schemes that, with appropriate beamforming, compensate for the orders-of-magnitude increase in free-space path-loss when compared to a sub-6 GHz environment (see \cite{Gao18,Busari18} and references therein). However, unlike sub-6 GHz communications, where processing is completely performed in the digital domain, at \gls{mmWave} frequencies the energy consumption and hardware cost associated to the use of a large number of antenna elements precludes this possibility. Instead, use is typically made of hybrid digital-analog architectures whereby a large number of antennas is interfaced through an analog front-end (implemented using phase shifters) to a much smaller number of \gls{RF} chains that take care of down-mixing and analog-to-digital conversion \cite{Heath16,Gao16}. The application of cell-free designs at \gls{mmWave} was first considered in \cite{Alonzo17} (subsequently expanded in \cite{Alonzo19}) in the context of uncorrelated channels and mainly targeting power allocation strategies for energy-efficiency maximization. The effects of fronthaul limitations in a cell-free massive \gls{MIMO} at \gls{mmWave} frequencies under correlated fading was studied in \cite{Femenias19} along with a proposal to conduct user selection in the likely event that the number of users in the system exceeds the number of \gls{RF} chains at the \glspl{AP}.

The deployment of a large number of antennas, either in a centralized form as in massive \gls{MIMO} or a distributed one as in cell-free massive \gls{MIMO}, raises concerns regarding energy consumption both from the point of view of energy efficiency but also when considering the overall emissions produced by the mobile communication industry. In particular, the whole of information and communication technology (ICT) industry has been estimated to contribute up to about 23\% of the global carbon footprint and about 51\% of global energy electricity consumption by 2030 \cite{Andrae15}. Addressing this issue, energy efficient wireless communication (also called \textit{green communication}) has been an important research thread for well over a decade and it is envisaged to continue to do so for the foreseeable future \cite{Gandotra17}. Among the many \textit{green} strategies that have been proposed, one that is specially effective in reducing the carbon footprint associated to cellular networks is the one based on the so-called \textit{switch on/off algorithms} (see, for instance, \cite{Wu15,Han16,Gandotra17} and references therein). Since most networks are designed and deployed to cope with fully loaded scenarios, a situation that most often is not sustained at all times, these techniques aim at dynamically turning on/off a fraction of the \glspl{BS} in response to variations in the user locations and traffic demands. Typically, these strategies are combined with specific forms of user association and cell zooming (i.e., cell breathing) see \cite{Tabassum14,Jia15,Xu17,Jiang18}) that boost their performance. Very recent research in \cite{femenias2020access} has shown that the application of \textit{switch on/off} algorithms in the context of cell-free massive \gls{MIMO}, whereby \glspl{AP} are dynamically (de)activated, has proved effective in optimizing the energy efficiency of the network. However, this work, which was framed in a sub-6 GHz context, relied on the assumption of having homogenously distributed users in the spatial domain, a condition seldom met in practice.

\subsection{Contributions}
Motivated by the above works and open issues, our main goal in this paper is to propose a green networking solution encompassing the design, performance evaluation and comparison of energy-efficient \gls{ASO} strategies for cell-free \gls{mmWave} massive \gls{MIMO} networks when considering non-uniform spatial traffic densities. In particular, the noteworthy contributions of our work can be summarized as follows:

\begin{itemize}[noitemsep,wide=0pt, leftmargin=\dimexpr\labelwidth + 2\labelsep\relax]

\item Tractable expressions are derived for the energy/spectral efficiencies, and the power consumption in cell-free \gls{mmWave} massive \gls{MIMO} networks for the particular case of \gls{ZF} precoding/combining for both the \gls{DL} and \gls{UL}. Remarkably, these mathematical expressions are able to account for the fact that there is a limited number of active \gls{RF} chains at each of the \glspl{AP} in the network.

\item In contrast to previous works on switch on/off algorithms for cell-free massive \gls{MIMO} networks (see, for instance, \cite{femenias2020access,van2020joint}), this study contemplates a realistic model to describe a non-uniform distribution of \glspl{MS}, thus capturing the heterogeneous nature of spatial traffic density in practical wireless networks \cite{lee2014spatial}.

\item Aside from recasting existing adaptive \gls{ASO} techniques to the scenario at hand, novel \gls{ASO} strategies are implemented that rely on statistical \gls{GoF} tests, and govern the process of (de)activating \glspl{AP} in such a way that the distribution of the resulting active \gls{AP} matches the non-uniform spatial distribution of \glspl{MS}. These \gls{GoF}-based strategies are shown to be computationally simple and to only depend on information regarding the location of \glspl{AP} and the long-term spatial distribution of \glspl{MS}. In comparison to previous techniques, our proposals attain a much better trade-off in terms of jointly assessing performance, complexity and implementability.

\item Extensive simulation results show the energy efficiency benefits of the proposed practical \gls{ASO} strategies under a comprehensive set of cell-free \gls{mmWave} massive \gls{MIMO} scenarios. In particular, the impact the number of \glspl{MS} in the network and the \gls{RF} infrastructure used at the \glspl{AP} have on the spectral/energy efficiency of the proposed network is evaluated under various non-uniform spatial distributions of \glspl{MS}.

\end{itemize}

\subsection{Paper organization and notational remarks}

The proposed green cell-free \gls{mmWave} massive \gls{MIMO} network is introduced in Section \ref{sec:System_model}, where different subsections are dedicated to describe the spatial modeling of the distribution of \glspl{MS}, the spatially correlated channel model at \gls{mmWave} bands, the \gls{RF} precoder/decoder design, the algorithm used to select the \glspl{MS} to beamform from each active \gls{AP}, the \gls{UL} training phase and, finally, the \gls{DL} and \gls{UL} payload transmission phases. The different performance metrics used in this paper, including the spectral efficiency, the power consumption model and the energy efficiency, are thoroughly evaluated in Section \ref{sec:Performance_metrics}. The proposed \gls{ASO} strategies, assuming scenarios with non-uniform spatial traffic distribution, are fully described in Section \ref{sec:ASO_strategies}. Numerical results and discussions are provided in Section \ref{sec:numerical_results} and, finally, Section \ref{sec:Conclusion} concludes the paper.

\textit{Notation}: Vectors and matrices are denoted by lower- and upper-case boldface symbols, respectively. The $q$-dimensional identity matrix is represented by $\boldsymbol{I}_q$. The operator $\|\bs{x}\|$ represents the Euclidian norm of vector $\bs{x}$, whereas $\bs{X}^{-1}$, $\bs{X}^T$, $\bs{X}^*$ and $\bs{X}^H$ denote the inverse, transpose, conjugate and conjugate transpose (also known as Hermitian) of matrix $\bs{X}$, respectively. With a slight abuse of notation, the operator $\diag(\bs{x})$ is used to denote a diagonal matrix with the entries of vector $\bs{x}$ on its main diagonal, whereas $\diag(\bs{X})$ is used to denote a vector containing the elements of the main diagonal of matrix $\bs{X}$. The expectation operator is denoted by $\mathbb{E}\{\cdot\}$. Finally, $\mathcal{CN}(\bs{m},\bs{R})$ denotes a complex Gaussian vector distribution with mean $\bs{m}$ and covariance matrix $\bs{R}$, $\mathcal{N}(0,\sigma^2)$ denotes a real valued zero-mean Gaussian random variable with standard deviation $\sigma$, and $\mathcal{U}[a,b]$ represents a random variable uniformly distributed in the range $[a,b]$. In order to ease the reading of this paper, Table \ref{tab:parameter_description} summarizes the definition of some of the most important parameters and variables used in the different subsections (see also Table \ref{tab:parameters} summarizing the default simulation parameters used in the numerical results section).

\begin{table}
  \renewcommand{\arraystretch}{1.15}
  \centering
  \caption{Summary of main parameters and variables}
  \begin{tabular}{p{1.75cm}p{5.8cm}}
    \hline
    \textbf{Parameter} & \textbf{Description} \\ \hline
    $M$ & Number of APs \\
    $M_A$ & Number of active APs \\
    $N$ & Number of antennas at each AP \\
    $L$ & Number of available RF chains at each AP \\
    $L$ & Number of aactive RF chains at each AP \\
    $K$ & Number of MSs \\
    $\check{\bs{h}}_{mk}$ & Channel between the $k$th MS and the $m$th AP \\
    $\overline{\bs{h}}_{mk}$ & Normalized LOS component of $\check{\bs{h}}_{mk}$ \\
    $\bs{h}_{mk}$ & Normalized NLOS component of $\check{\bs{h}}_{mk}$ \\
    $\check{\bf{R}}_{mk}$ & Spatial covariance matrix of $\check{\bs{h}}_{mk}$ \\
    $\bf{R}_{mk}$ & Spatial covariance matrix of $\bs{h}_{mk}$ \\
    $\bs{W}_m^{\text{RF}}$ & Analog RF precoder/combiner stage at the $m$th AP \\
    $\bs{W}_{d\,m}^{\text{BB}}$/$\bs{W}_{u\,m}^{\text{BB}}$ & Digital precoder/combiner stage at the $m$th AP \\
    $\bs{g}_{mk}$ & Equivalent channel (including RF precoder/decoder) between the $k$th MS and the $m$th AP \\
    $\hat{\bs{g}}_{mk}$ & MMSE estimation of $\bs{g}_{mk}$ \\
    $\check{\bs{H}}_m$ & MIMO channel between the $m$th AP and the $K$ MSs \\
    $\bs{G}_m$ & Equivalent MIMO channel (including RF precoder/decoder) between the $m$th AP and the $K$ MSs \\
    ${S_e}_d(\bs{\upsilon})$/${S_e}_u(\bs{\omega})$ & DL/UL spectral efficiency as a function of the vector of DL/UL \\
    ${P_T}_d(\bs{\upsilon})$/${P_T}_u(\bs{\omega})$ & DL/UL power consumption as a function of the vector of DL/UL power control coefficients \\
    ${E_e}_d(\bs{\upsilon})$/${E_e}_u(\bs{\omega})$ & DL/UL energy efficiency as a function of the vector of DL/UL power control coefficients \\
    ${E_e}(\bs{\upsilon},\bs{\omega})$ & Weighted energy efficiency as a function of the vector of DL and UL power control coefficients \\
    \hline
  \end{tabular}
  \label{tab:parameter_description}
\end{table}

\begin{figure}[t]
    \centering
    \begin{subfigure}[t]{\linewidth}
        \centering
        \includegraphics[width=0.9\linewidth]{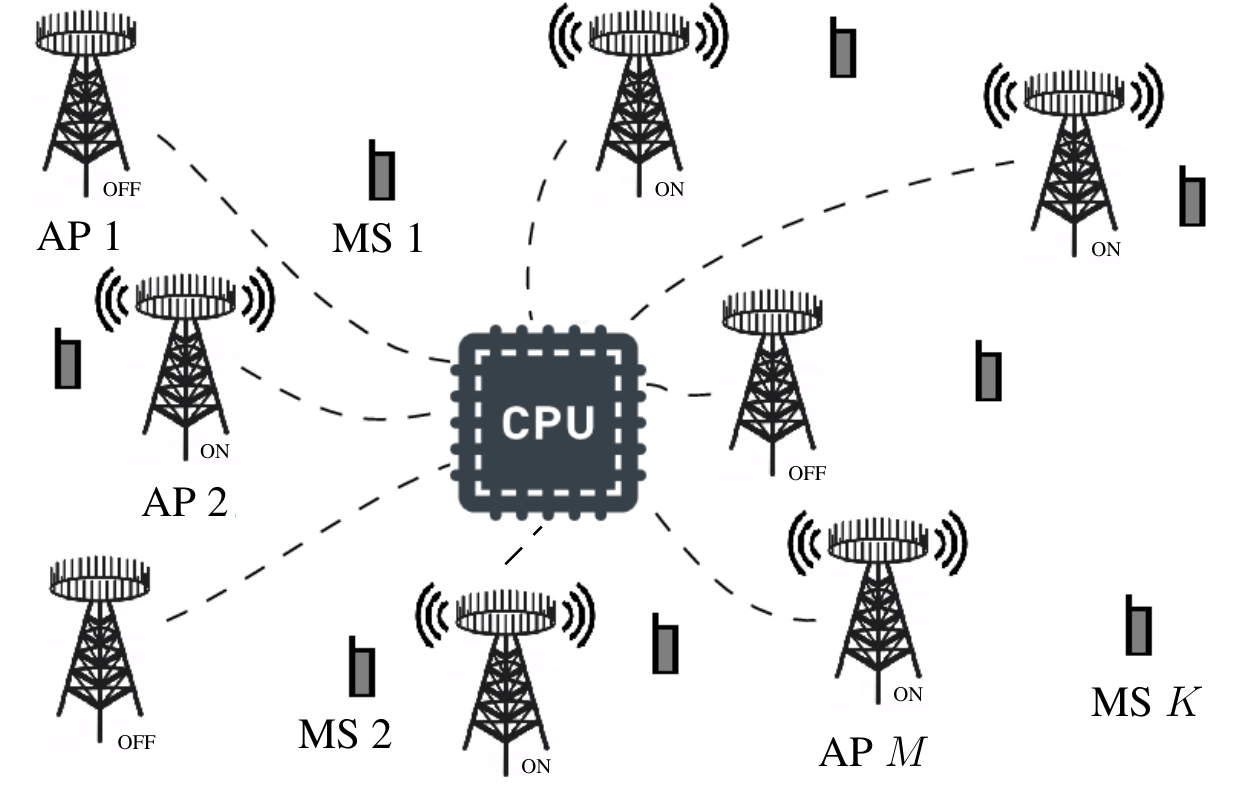}
        \caption{Cell-free \gls{mmWave} massive \gls{MIMO} network where $M_A$ active \glspl{AP} (out of $M$ available \glspl{AP}) provide service to $K$ \glspl{MS}.}
        \label{fig:network}
    \end{subfigure}%

    \begin{subfigure}[t]{\linewidth}
        \centering
        \includegraphics[width=0.95\linewidth]{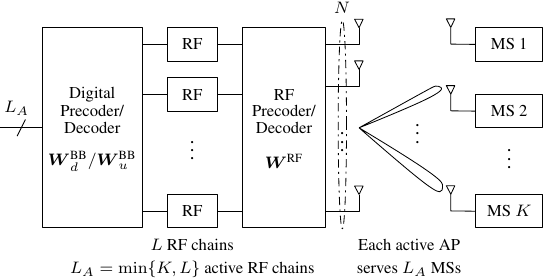}
        \caption{Generic transmitter/receiver architecture: hybrid analog-digital precoder/decoder with $L$ \gls{RF} chains, and fully connected phase shifter network.}
        \label{fig:architecture}
    \end{subfigure}
    \caption{System model of a cell-free \gls{mmWave} massive \gls{MIMO} network using \gls{ASO} strategies and hybrid analog-digital precoding/decoding schemes.}
    \label{fig:system_model}
\end{figure}

\section{System model}
\label{sec:System_model}

As shown in Fig. \ref{fig:network}, we consider a cell-free \gls{mmWave} massive \gls{MIMO} network where $M$ randomly distributed \glspl{AP}, each equipped with an array of $N$ antennas and connected to a \gls{CPU} via an infinite-capacity error-free fronthaul link, can be activated to provide service to $K$ single-antenna \glspl{MS}. The process of activation/deactivation of \glspl{AP} is basically driven by the implementation of \gls{ASO} strategies, with \glspl{AP} in active (ON) and sleep (OFF) modes being indexed by the sets $\mathcal{M}^A=\{m_1^A,\ldots,m_{M_A}^A\}$ and $\mathcal{M}^S=\{m_1^S,\ldots,m_{M_S}^S\}$, respectively, where $\mathcal{M}^A\cap\mathcal{M}^S=\emptyset$ and $\mathcal{M}^A\cup\mathcal{M}^S=\{1,\ldots,M\}$. Moreover, as the implementation of a dedicated \gls{RF} chain to each antenna results in unaffordable energy consumption and hardware cost in \gls{mmWave}-based massive \gls{MIMO} systems, the number of required \gls{RF} chains is reduced by relying on the use of hybrid analog-digital precoding schemes. In particular, as shown in Fig. \ref{fig:architecture}, it is assumed in this paper that each \gls{AP} is equipped with $L \leq N$ \gls{RF} chains and that a fully-connected architecture is arranged where each \gls{RF} chain is connected to all the antenna elements using $N$ analog phase shifters.

The communication between the active \glspl{AP} and the \glspl{MS} is coordinated by the \gls{CPU} through the use of a half-duplex \gls{TDD} algorithm in which each frame is divided into three phases, namely, the \gls{UL} training phase, the \gls{UL} payload data transmission phase and the \gls{DL} payload data transmission phase. During the \gls{UL} training phase, all \glspl{MS} transmit training pilots allowing the active \glspl{AP} to estimate the propagation channels to every \gls{MS} in the network. Channel estimates are then used to detect the signals transmitted from the \glspl{MS} in the \gls{UL} payload data transmission phase and to compute the precoding filters governing the \gls{DL} payload data transmission. The combined duration/bandwidth of the training, \gls{UL} and \gls{DL} phases, denoted as $\tau_p$, $\tau_u$ and $\tau_d$, respectively, should not exceed the coherence time/bandwidth of the channel, denoted as $\tau_c$, that is, $\tau_p+\tau_d+\tau_u\leq \tau_c$, with all these intervals specified in samples (or channel uses) on a time-frequency grid. It is worth pointing out at this point that although the small-scale parameters characterizing the propagation channels linking the \glspl{AP} and \glspl{MS} can only be safely assumed to be static over a coherence time-frequency interval of $\tau_c$ samples, the large-scale parameters (i.e., path loss propagation losses and spatial covariance matrices) can be safely assumed to be static over a time-frequency interval ${\tau_L}_c \gg \tau_c$ \cite{Marzetta16,Heath16}. These particular channel characteristics will be leveraged in the next subsections to simplify both the channel estimation and the precoding/combining processes.

\subsection{Spatial modeling of the \gls{MS} distribution}
\label{subsec:Spatial_modeling}

As we are interested in exploring the impact \gls{ASO} strategies may have on the performance of cell-free \gls{mmWave} massive \gls{MIMO} networks with a non-uniform spatial traffic distribution, the location of \glspl{MS} on the service area will be modeled using the approach proposed by Lee \textit{et al.} in \cite{lee2014spatial}. This spatial traffic model generates large-scale spatial traffic variations by resorting to the use of sums of sinusoids capturing the characteristics of spatially correlated log-normally distributed traffic. In particular, let us consider an square area $\mathcal{S}$ of side $D$. This target region is tessellated in a regular grid of $N_X$ by $N_Y$ rectangular cells (or pixels). A cell (or pixel) $(x,y)$ where $x\in\{1,\ldots,N_X\}$ and $y\in\{1,\ldots,N_Y\}$, is characterized by a traffic density demand $\rho_{x,y}$ (in \glspl{MS} per pixel). To generate a log-normal distributed traffic map, a Gaussian random field is first produced as
\begin{equation}
\begin{split}
   \rho^G_{x,y} = \frac{2}{\sqrt{T}} \sum_{t=1}^T &\cos\left(i^{(\text{u})}_t \Re\{p_{x,y}\} + \theta^{(\text{u})}_t\right)\\
                                                  &\times \cos\left(j^{(\text{u})}_t \Im\{p_{x,y}\} + \phi^{(\text{u})}_t\right),
\end{split}
\end{equation}
where $p_{x,y}=\Re\{p_{x,y}\}+j\Im\{p_{x,y}\}$ is used to denote the location (on a complex plane) of the center of pixel $(x,y)$. The angular frequencies $i^{(\text{u})}_t$ and $j^{(\text{u})}_t$ are random variables uniformly distributed in the range $[0, \omega^{\mathcal{S}}_{\max}]$, where $\omega^{\mathcal{S}}_{\max}$ is defined as the maximum spatial spread used to control the rate of fluctuations of the random field in the area $\mathcal{S}$, and phases $\theta^{(\text{u})}_t$ and $\phi^{(\text{u})}_t$ are random variables uniformly distributed in the range $[0, 2 \pi]$. According to the central limit theorem, for a large enough value of $T$, the symbol $\rho^G_{x,y}$ can be approximated as a standard Gaussian random variable. In \cite{lee2014spatial}, Lee \textit{et al.} found that using $T=10$ provides sufficiently accurate results.

\begin{figure}
  \centering
  \includegraphics[width=7.8cm]{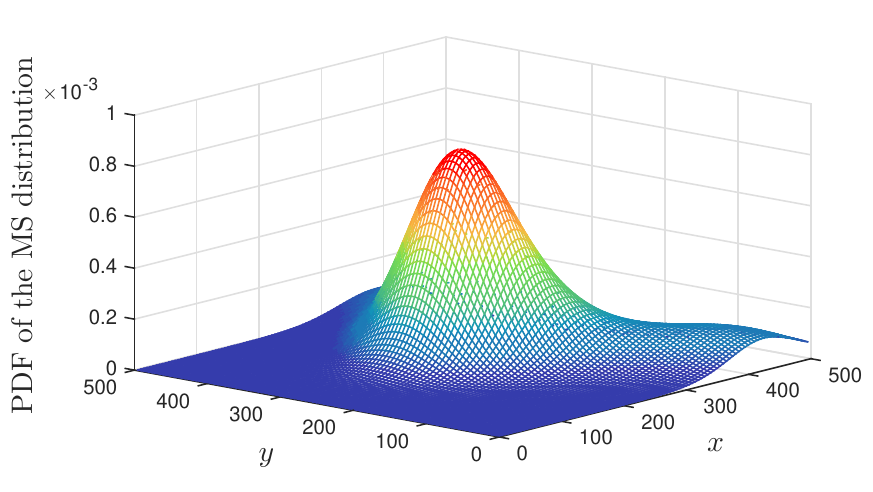}
  \caption{Probability density function of the number of \glspl{MS} per pixel on a square grid of side $D=500$ m and square pixels of side $5$ m in an urban scenario (see parameters in \cite[Table I]{lee2014spatial}).}
  \label{fig:MS_pdf}
\end{figure}

By calculating the exponential function of $\rho^G_{x,y}$ with the location parameter $\mu^{\mathcal{S}}$ and the scaling parameter $\sigma^{\mathcal{S}}$, a spatial traffic density matrix can be obtained whose elements $\rho_{x,y}$, for all $x\in\{1,\ldots,N_X\}$ and $y\in\{1,\ldots,N_Y\}$, can be expressed as
\begin{equation}
   \rho_{x,y}=\exp\left(\sigma^{\mathcal{S}} \rho^G_{x,y} + \mu^{\mathcal{S}}\right).
\end{equation}
These random variables are log-normally distributed and by controlling the parameters $\mu^{\mathcal{S}}$ and $\sigma^{\mathcal{S}}$ the corresponding log-normal distribution can be scaled to fit the statistics of the traffic distribution experienced in different scenarios \cite{lee2014spatial}. The \gls{pdf} of the number of \glspl{MS} per pixel can be finally calculated as
\begin{equation}
   f^{\gls{MS}}_{x,y}=\frac{\rho_{x,y}}{\sum_{x'=1}^{N_X} \sum_{y'=1}^{N_Y} \rho_{x',y'}}.
   \label{eq:MS_pdf}
\end{equation}
For instance, Fig. \ref{fig:MS_pdf} represents the \gls{pdf} characterizing the number of \glspl{MS} per pixel on a square grid of side $D=500$ m and square pixels of side $5$ m using $\omega^{\mathcal{S}}_{\max}=$0.012673, $\mu^{\mathcal{S}}=$17.7956 and $\sigma^{\mathcal{S}}=$2.1188, which are characteristic values of urban scenarios according to \cite[Table I]{lee2014spatial}.

\subsection{Channel model}

Compared to lower frequency bands, propagation in the \gls{mmWave} band is characterized by very high distance- and penetration-based propagation losses that lead to sparse scattering multipath propagation, thus boosting the importance of \gls{LOS} propagation, reflection and blockage. Furthermore, high antenna correlation levels may be expected because of the use of mmWave transmitters and receivers with tightly-packet large antenna arrays.
The peculiarities of these mmWave channels can be captured by using a simplified clustered channel model version inspired by the \gls{3GPP} for Urban Micro-cell scenarios described in \cite{3GPPmmWaveR14} and by a variety of research works (see \cite{Akdeniz14,Samimi14,Neil17,Femenias19} and references therein).

According to this simplified clustered channel model, the propagation link between the $m$th \gls{AP} and \gls{MS} $k$ can be either in outage, in \gls{LOS} or in \gls{NLOS} conditions with probabilities \cite{Akdeniz14}
\begin{subequations}
\begin{equation}
   p_{\text{out}}(d_{mk})=\max\left(0,1-e^{-a_{\text{out}}d_{mk}+b_{\text{out}}}\right),
\end{equation}
\begin{equation}
   p_{\text{LOS}}(d_{mk})=\left(1-p_{\text{out}}(d_{mk})\right) e^{-a_{\text{LOS}}d_{mk}},
\end{equation}
\begin{equation}
   p_{\text{NLOS}}(d_{mk})=1-p_{\text{out}}(d_{mk})-p_{\text{LOS}}(d_{mk}),
\end{equation}
\end{subequations}
respectively, where $d_{mk}$ is the distance (in meters) between the \gls{AP} and the \gls{MS}. The parameters governing these probabilities are set to $1/a_{\text{out}}=30$~m, $b_{\text{out}}=5.2$, and $1/a_{\text{LOS}}=67.1$~m (see \cite[Table I]{Akdeniz14}). When in outage conditions, this propagation link will be characterized by infinite propagation losses. When in \gls{LOS} or \gls{NLOS} conditions, however, a standard linear model with shadowing will be used that can be expressed as \cite{3GPPmmWaveR14}
\begin{equation}
   L_{mk}[dB]=\alpha+10 \beta \log_{10}(d_{mk})+\chi_{mk},
\end{equation}
where $\alpha$ and $\beta$ are frequency-dependent least square fits of floating intercept and slope, respectively, and are selected according to whether the link is in \gls{LOS} or \gls{NLOS}, and the large-scale shadow fading component $\chi_{mk}$ is modelled as a zero mean spatially correlated normal random variable with standard deviation $\sigma_\chi$.

Based on the large-scale propagation loss model just described, the channel vector $\breve{\bs{h}}_{mk} \in \mathbb{C}^{N \times 1}$ from the $k$th \gls{MS} to the $m$th \gls{AP} (including both large-scale and small-scale fading) can be generically characterized as a Ricean fading channel consisting of a \gls{LOS} component on top of a Rayleigh distributed component modelling the scattered multipath. That is,
\begin{equation}
   \breve{\bs{h}}_{mk}=\sqrt{\frac{K_{mk}}{K_{mk}+1}}\overline{\bs{h}}_{mk}+\sqrt{\frac{1}{K_{mk}+1}}\bs{h}_{mk},
\end{equation}
with a normalized \gls{LOS} component
\begin{equation}
   \overline{\bs{h}}_{mk}=\alpha_{mk} \bs{a}\left(\overline{\theta}_{mk,1},\overline{\phi}_{mk,1}\right),
\end{equation}
and a normalized \gls{NLOS} component
\begin{equation}
\begin{split}
   \bs{h}_{mk}=&\sum_{c=1}^{C_{mk}}\sum_{p=1}^{P_{mk}} \alpha_{mk,cp} \bs{a}\left(\theta_{mk,cp},\phi_{mk,cp}\right),
\end{split}
\end{equation}
where $K_{mk}$ is the Ricean $K$-factor, with $K_{mk}=0$ for \gls{NLOS} propagation links and $10 \log_{10}(K_{mk}) \sim \mathcal{N}\left(\mu_K,\sigma_K^2\right)$ for \gls{LOS} propagation links. The parameter $\alpha_{mk}=10^{-L_{mk}/20} e^{j\kappa_{mk}}$, with $\kappa_{mk} \sim \mathcal{U}[0, 2\pi]$, is used to denote the large-scale complex channel gain of the \gls{LOS} component, $C_{mk}$ and $P_{mk}$ are the number of contributing scattering clusters of the \gls{NLOS} component and the number of propagation paths per cluster, respectively, $\alpha_{mk,cp}$ is the complex small-scale fading gain on the $p$th path of cluster $c$, and $\bs{a}\left(\theta_{mk,cp},\phi_{mk,cp}\right)$ is the normalized array response vector of the \gls{AP} at the azimuth and elevation angles $\theta_{mk,cp}$ and $\phi_{mk,cp}$, respectively.

As suggested by Akdeniz \textit{et al.} in \cite{Akdeniz14}, $\theta_{mk,cp}$ can be generated as a wrapped Gaussian around the cluster central angle $\overline{\theta}_{mk,c}$ with standard deviation given by the \gls{rms} azimuth angular spreads for the cluster. Furthermore, $\phi_{mk,cp}$ can be generated as a Laplacian around the cluster central angle $\overline{\phi}_{mk,c}$ with scale parameters given by the \gls{rms} elevation angular spreads for the cluster. The azimuth cluster central angle $\overline{\theta}_{mk,c}$ is uniformly distributed in the range $[-\pi,\pi]$ and the elevation cluster central angle $\overline{\phi}_{mk,c}$ is set to the corresponding \gls{LOS} elevation angle. The cluster \gls{rms} angular spreads are exponentially distributed with a mean equal to $1/\lambda_{\gls{rms}}$ that depends on whether we are considering the azimuth or elevation directions. The small-scale scattering fading gains are distributed as
\begin{equation}
   \alpha_{mk,cp} \sim \mathcal{CN}\left(0,\gamma_{mk,c}10^{-L_{mk}/10}\right),
\end{equation}
where the cluster $c$ is assumed to contribute to the scatter fading with a fraction of power given by
\begin{equation}
   \gamma_{mk,c}=\frac{N \gamma'_{mk,c}}{P_{mk}\sum_{j=1}^{C_{mk}} \gamma'_{mk,j}},
\end{equation}
with
\begin{equation}
   \gamma'_{mk,j}=U_{mk,j}^{r_\tau-1} 10^{Z_{mk,j}/10},
\end{equation}
where $U_{mk,j} \sim \mathcal{U}[0,1]$, $Z_{mk,j} \sim \mathcal{N}(0,\zeta^2)$, and the constants $r_\tau$ and $\zeta^2$ being treated as model parameters \cite{Akdeniz14}.

Using this channel propagation model, the spatial covariance matrix of the scattered multipath component $\bs{h}_{mk}$ can be obtained as
\begin{equation}
\begin{split}
   \bs{R}_{mk} &=\mathbb{E}\left\{\bs{h}_{mk} \bs{h}_{mk}^H \right\} \\
   &= 10^{-L_{mk}/10}\sum_{c=1}^{C_{mk}}\gamma_{mk,c} \\
   & \times \sum_{p=1}^{P_{mk}} \bs{a}\left(\theta_{mk,cp},\phi_{mk,cp}\right)\left(\bs{a}\left(\theta_{mk,cp},\phi_{mk,cp}\right)\right)^H.
\end{split}
\end{equation}
and the spatial covariance matrix of the resulting channel vector $\breve{\bs{h}}_{mk}$ can thus be expressed as
\begin{equation}
\begin{split}
   \breve{\bs{R}}_{mk} &=\mathbb{E}\left\{\breve{\bs{h}}_{mk} \breve{\bs{h}}_{mk}^H \right\}\\
   &= \frac{K_{mk}}{K_{mk}+1}\overline{\bs{h}}_{mk}\overline{\bs{h}}_{mk}^H + \frac{\bs{R}_{mk}}{K_{mk}+1}.
\end{split}
\end{equation}
As stated by \"Ozdogan \emph{et al.} in \cite{Ozdogan19}, these spatial covariance matrices, as well as the propagation path losses, Ricean $K$-factors, and channel means, can be considered to be constant over frequency-time intervals much larger than the coherence interval $\tau_c$ and, consequently, they can be straightforwardly estimated in practice using the sample mean and sample covariance matrices  \cite{Bjornson17,Haghighatshoar17a,Neumann18,Upadhya18}.

\subsection{RF precoder/combiner design}

Without loss of essential generality, it is assumed in this paper that a hybrid precoding technique is used in which each \gls{RF} chain is dedicated to one and only one \gls{MS}. In particular, if $K\leq L$, all active \glspl{AP} in the network provide service to the $K$ \glspl{MS} and only $K$ \gls{RF} chains per \gls{AP} are activated (one for each \gls{MS} in the network). If $K > L$, instead, each active \gls{AP} can only serve a subset of $L$ \glspl{MS} (one for each \gls{RF} chain) and thus, an algorithm must be devised to decide which are the subsets of \glspl{MS} to be beamformed from each of the active \glspl{AP} while ensuring that the $K$ \glspl{MS} are simultaneously served. The number of active \gls{RF} chains per \gls{AP} can then be generically expressed as $L_A=\min\{K,L\}$ (see Fig. \ref{fig:architecture}).

Channel reciprocity is exploited by implementing an $N \times L_A$ \gls{RF} precoding matrix $\bs{W}_m^{\text{RF}}$, describing the effects of the active analog phase shifters at the $m$th active \gls{AP}, which is common to the \gls{UL} (\gls{RF} combining phase) and \gls{DL} (\gls{RF} precoding phase). Furthermore, denoting by $\mathcal{K}_m=\left\{\kappa_{m 1}, \ldots, \kappa_{m L_A}\right\}$ the set of $L_A$ \glspl{MS} beamformed by the $m$th \gls{AP}, it is assumed that $\bs{W}_m^{\text{RF}}$ is a function of only the spatial channel covariance matrices $\left\{\breve{\bs{R}}_{mk}\right\}_{k\in\mathcal{K}_m}$, known at the $m$th \gls{AP} through spatial channel covariance estimation for hybrid analog-digital \gls{MIMO} precoding architectures \cite{Adhikary14,Mendez15,Park17TSC}. The use of long-term channel statistics such as the spatial covariance matrices is a reasonable approach as they vary over very long time scales and, moreover, they can be safely assumed to be uniform across the whole system bandwidth, thus providing a good solution to the problem of designing a common analog precoder for all subcarriers \cite{Park17TSC}.

The Hermitian covariance matrix of the propagation channel linking \gls{MS} $k$ and \gls{AP} $m$ can be factorized using eigen-decomposition as $\breve{\bs{R}}_{mk}=\bs{U}_{mk}\bs{\Lambda}_{mk}\bs{U}_{mk}^H$, where $\bs{\Lambda}_{mk}=\diag\left(\left[\lambda_{mk,1}\,\ldots\,\lambda_{mk,r_{mk}}\right]\right)$ contains the $r_{mk}$ non-null eigenvalues of $\breve{\bs{R}}_{mk}$, and $\bs{U}_{mk}$ is the $N \times r_{mk}$ matrix of the corresponding eigenvectors. This eigen-factorization can be exploited to design the analog \gls{RF} precoder/combiner stage by using the well-known (constrained) statistical eigen beamforming \cite{Park17TSC,Mai18}, where
\begin{equation}
\begin{split}
   \bs{W}_m^{\text{RF}}=&\begin{bmatrix}
                    \bs{w}_{m \kappa_{m 1}}^{\text{RF}} & \ldots & \bs{w}_{m \kappa_{m L_A}}^{\text{RF}}
                 \end{bmatrix}\\
                =&\begin{bmatrix}
                    e^{-j\angle\bs{u}_{m \kappa_{m 1},\max}} & \ldots & e^{-j\angle\bs{u}_{m \kappa_{m L_A},\max}}
                 \end{bmatrix},
\end{split}
\end{equation}
with $\bs{u}_{mk,\max}$ denoting the dominant eigenvector of $\breve{\bs{R}}_{mk}$ associated to the maximum eigenvalue $\lambda_{mk,\max}$, and the function $\angle\bs{x}$ returning the phase angles, in radians, for each element of the complex vector $\bs{x}$. The equivalent channel vector between \gls{MS} $k$ and \gls{AP} $m$, including the analog \gls{RF} precoder/combiner, can then be defined as
\begin{equation}
   \bs{g}_{mk} = {\bs{W}_m^{\text{RF}}}^T \breve{\bs{h}}_{mk}\ \in\ \mathbb{C}^{L_A \times 1},
\end{equation}
whose dimension $L_A$ is typically much less than the number of antennas of the massive \gls{MIMO} array, thus making the small-scale training phase computationally simpler.

\begin{algorithm}[t]
\caption{Selection of \glspl{MS} to beamform from each \gls{AP}}
\label{alg:MS selection}
\begin{algorithmic}
\STATE \textbf{Input:} $\xi_{mk}\,\forall mk, \mathcal{M}^A$, $M_A$, $L$, $K$

\STATE \textbf{Initialization:} $\mathcal{K}_m^{(0)}=\{1,\ldots,K\}\ \forall m$

\STATE $\qquad\qquad\qquad\mathcal{M}_k^{(0)}=\mathcal{M}^A\ \forall k$

\STATE $\qquad\qquad\qquad\mathcal{A}=\mathcal{M}^A$

\FOR {$i=1:M_A(K-L)$}


    \STATE $(m^*,k^*)=\arg\displaystyle\max_{k\in\{1,\ldots,K\}}\min_{m\in\mathcal{A}}\sum_{n\in\mathcal{M}_k^{(i-1)}\setminus m} \xi_{nk}$

    \STATE \{Remove $k^*$ from the set of MSs served by AP $m^*$\}

    \STATE $\mathcal{K}_{m^*}^{(i)}=\mathcal{K}_{m^*}^{(i-1)}\setminus k^*$

    \STATE \{Remove $m^*$ from the sets of APs serving MS $k^*$\}

    \STATE $\mathcal{M}_{k^*}^{(i)}=\mathcal{M}_{k^*}^{(i-1)}\setminus m^*$

    \STATE \{Update set of APs already serving $L$ MSs\}

    \IF {$\left|\mathcal{K}_{m^*}^{(i)}\right|=L$}
        \STATE $\mathcal{A}=\mathcal{A}\setminus m^*$
    \ENDIF
\ENDFOR
\STATE \textbf{Output:} $\mathcal{K}_m=\mathcal{K}_m^{(M_A(K-L))}\,\forall m \in \mathcal{M}^A$
\end{algorithmic}
\end{algorithm}

\subsection{Selection of \glspl{MS} to beamform from each \gls{AP}}

As we only consider the use of analog precoding/decoding stages in which each \gls{RF} chain at a given \gls{AP} is dedicated to a single \gls{MS}, in those cases in which the number of \glspl{MS} is greater than the number of available \gls{RF} chains at each \gls{AP} (i.e., $K > L$), the group of $L$ \glspl{MS} to beamform from each active \gls{AP} in the cell-free network, indexed by the sets  $\mathcal{K}_m=\left\{\kappa_{m1},\ldots,\kappa_{mL}\right\}$, for all $m\in\{1,\ldots,M\}$, will have to be selected. Furthermore, as the \gls{RF} beamforming/decoding matrices at the \glspl{AP} are designed assuming only the availability of the large-scale spatial channel covariance matrices, this selection process can only be based on this large-scale \gls{CSI}. Inspired by the Frobenius norm-based suboptimal user selection algorithm proposed by Shen \textit{et al.} in \cite{Shen06}, an iterative selection algorithm was proposed in \cite{Femenias19} that, under the constraint that each \gls{AP} can only beamform to $L$ \glspl{MS}, aims at maximizing the minimum average sum energy of the equivalent channels between the $M_A$ active \glspl{AP} and any of the $K$ \glspl{MS} in the network. Note that, using this algorithm, each active \gls{AP} will beamform to exactly $L$ \glspl{MS} and each \gls{MS} will be beamformed by at least one \gls{AP}.

At the beginning of the $i$th iteration of the algorithm, a simple edge-weighted directed graph with $M_A$ source nodes and $K$ sink nodes is used to represent the cell-free massive \gls{MIMO} network. In this directed graph, the $m$th source node, which represents the $m$th active \gls{AP}, is connected to a group $\mathcal{K}_m^{(i)}$ of sink nodes, used to represent the \glspl{MS} to be \textit{potentially} beamformed from the $m$th active \gls{AP}. The average energy of the equivalent channel linking the $m$th active \gls{AP} and \gls{MS} $l\in \mathcal{K}_m^{(i)}$, which can be obtained as
\begin{equation}
   \xi_{ml}=\mathbb{E}\left\{\left|{\bs{w}_{ml}^{\text{RF}}}^T\breve{\bs{h}}_{ml}\right|^2\right\} = {\bs{w}_{ml}^{\text{RF}}}^T \breve{\bs{R}}_{ml} {\bs{w}_{ml}^{\text{RF}}}^*,
\end{equation}
is used to weight the connection (edge) joining the $m$th source node and the $l$th sink node in $\mathcal{K}_m^{(i)}$. Using this notation, the average sum energy of the equivalent channels between the $M_A$ active \glspl{AP} and \gls{MS} $k$ at the beginning of the $i$th iteration can be obtained as
\begin{equation}
   \mathcal{E}_k^{(i)} = \sum_{m\in\mathcal{M}_k^{(i)}} \xi_{mk},
\end{equation}
where $\mathcal{M}_k^{(i)}$ is the set of active \glspl{AP} selected in previous iterations to beamform to \gls{MS} $k$. The reverse-delete algorithm is used in this iteration to remove the edge (i.e., the \gls{RF} chain and associated beamformer) coming from one of those active \glspl{AP} still beamforming to more than $L$ \glspl{MS} that maximizes the minimum average sum energy per \gls{MS} after removal. The proposed algorithm, starting with a fully connected graph, stops after $M_A(K-L)$ iterations with all active \glspl{AP} in $\mathcal{M}^A$ beamforming to exactly $L$ \glspl{MS}. A mathematical pseudocode for this algorithm is shown in Algorithm \ref{alg:MS selection}.

\subsection{Small-scale training phase: Channel estimation}

During the \gls{UL} training phase, all $K$ \glspl{MS} simultaneously transmit pilot sequences of $\tau_p$ samples to the active \glspl{AP} and thus, the $L_A \times \tau_p$ received \gls{UL} signal matrix at the $m$th active \gls{AP} can be expressed as
\begin{equation}
   {\bs{Y}_p}_m=\sqrt{\tau_p P_p}\sum_{k'=1}^K \bs{g}_{mk'} \bs{\varphi}_{k'}^T+{\bs{N}_p}_m,
\end{equation}
where $P_p$ is the transmit power of each pilot symbol, $\bs{\varphi}_k$ denotes the $\tau_p\times 1$ pilot signal allocated to \gls{MS} $k$, with $\|\bs{\varphi}_k\|^2=1$, and ${\bs{N}_p}_m$ is an $L_A\times\tau_p$ matrix of i.i.d. additive noise samples with each entry distributed as\footnote{Note that in the \gls{UL} of a fully-connected hybrid beamforming architecture each reception chain is composed of $N$ antenna elements, each connected to a low-noise amplifier (LNA) characterized by a power gain $G_{\textrm{LNA}}$ and a noise temperature $T_{\text{LNA}}$. Each of the $N$ LNAs feeds an analog passive phase shifter characterized by an insertion loss $L_{\text{PS}}$. The outputs of the $N$ phase shifters are introduced to a power combiner whose insertion losses are typically proportional to the number of inputs, that is, $L_{\text{PC}}=N L_{\text{PC}_{in}}$. Finally, the output of the power combiner is introduced to an $\gls{RF}$ chain characterized by a power gain $G_{\text{RF}}$ and a noise temperature $T_{\text{RF}}$. Thus, the equivalent noise temperature of each receive chain can be obtained as $T_u=N \left(T_0+T_{\text{LNA}}+\frac{T_0(L_{\text{PS}}L_{\text{PC}_{in}}-1)}{G_{\text{LNA}}}+\frac{T_{\text{RF}}L_{\text{PS}}L_{\text{PC}_{in}}}{G_{\text{LNA}}}\right)$.}  $\mathcal{CN}(0,\sigma_u^2(N))$. Note that, since in most practical scenarios it holds that $K>\tau_p$, a given pilot sequence will be allocated to more than one \gls{MS} and, hence, pilot contamination will arise \cite{Marzetta16,Elijah16}.

As previously stated, considering scenarios where \glspl{MS} move slowly, it is reasonable to assume that the Ricean $K$-factors $K_{mk}$, the \gls{LOS} components $\overline{\bs{h}}_{mk}$, and the scatter fading correlation matrices $\bs{R}_{mk}$ change slowly and can be perfectly known at the $m$th active \gls{AP}, for all $k$ \cite{Ngo18Rice}. Under this assumption, we can define
\begin{equation}
\begin{split}
    &\breve{\bs{y}}_{p\,mk}=\left({\bs{Y}_p}_m-\mathbb{E}\left\{{\bs{Y}_p}_m\right\}\right)\bs{\varphi}_k^* \\
    &\quad=\left(\sum_{k'=1}^K \sqrt{\frac{\tau_p P_p}{K_{mk'}+1}}{\bs{W}_m^{\text{RF}}}^T\bs{h}_{mk'}\bs{\varphi}_{k'}^T+{\bs{N}_p}_m\right)\bs{\varphi}_k^*
\end{split}
\end{equation}
and
\begin{equation}
   \breve{\bs{g}}_{mk}=\bs{g}_{mk}-\mathbb{E}\left\{\bs{g}_{mk}\right\}=\sqrt{\frac{1}{K_{mk}+1}}{\bs{W}_m^{\text{RF}}}^T\bs{h}_{mk},
\end{equation}
and then derive the \gls{MMSE} estimate for the channel between the $k$th \gls{MS} and the $m$th active \gls{AP} as \cite{Kay93,Ngo18Rice}
\begin{equation}
\begin{split}
   \hat{\bs{g}}_{mk}&=\sqrt{\frac{K_{mk}}{K_{mk}+1}}{\bs{W}_m^{\text{RF}}}^T\overline{\bs{h}}_{mk}\\ &+\mathbb{E}\left\{\breve{\bs{y}}_{p\,mk}\breve{\bs{g}}_{mk}^H\right\}\left(\mathbb{E}\left\{\breve{\bs{y}}_{p\,mk}\breve{\bs{y}}_{p\,mk}^H\right\}\right)^{-1}\breve{\bs{y}}_{p\,mk}\\
   &=\sqrt{\frac{K_{mk}}{K_{mk}+1}}{\bs{W}_m^{\text{RF}}}^T\overline{\bs{h}}_{mk}\\
   &+\frac{\sqrt{\tau_p P_p}}{K_{mk}+1} \bs{R}_{mk}^{\text{RF}} \bs{\Psi}_{mk}^{-1} \breve{\bs{y}}_{p\,mk},
\end{split}
\end{equation}
where
\begin{equation}
   \bs{R}_{mk}^{\text{RF}}={\bs{W}_m^{\text{RF}}}^T \bs{R}_{mk} {\bs{W}_m^{\text{RF}}}^*,
\end{equation}
and
\begin{equation}
\begin{split}
   \bs{\Psi}_{mk}&=\tau_p P_p \sum_{k'=1}^K \frac{\bs{R}_{mk'}^{\text{RF}}}{K_{mk'}+1} \left|\bs{\varphi}_{k'}^H\bs{\varphi}_k\right|^2 + \sigma_u^2(N) \bs{I}_{L_A}.
\end{split}
\end{equation}
The channel estimate $\hat{\bs{g}}_{mk}$ and the \gls{MMSE} channel estimation error $\tilde{\bs{g}}_{mk}=\bs{g}_{mk}-\hat{\bs{g}}_{mk}$ are uncorrelated random vectors distributed as
\begin{equation}
   \hat{\bs{g}}_{mk} \sim \mathcal{CN}\left(\sqrt{\frac{K_{mk}}{K_{mk}+1}}{\bs{W}_m^{\text{RF}}}^T\overline{\bs{h}}_{mk},\hat{\bs{A}}_{mk} \right),
\end{equation}
and $\tilde{\bs{g}}_{mk} \sim \mathcal{CN}\left(\bs{0},\tilde{\bs{A}}_{mk}\right)$, respectively, where
\begin{equation}
    \hat{\bs{A}}_{mk}=\frac{\tau_p P_p \bs{R}_{mk}^{\text{RF}} \bs{\Psi}_{mk}^{-1} \left(\bs{R}_{mk}^{\text{RF}}\right)^H}{\left(K_{mk}+1\right)^2},
\end{equation}
is the covariance matrix of $\hat{\bs{g}}_{mk}$ and
\begin{equation}
\begin{split}
    \tilde{\bs{A}}_{mk}&=\mathbb{E}\left\{\tilde{\bs{g}}_{mk}\tilde{\bs{g}}_{mk}^H\right\}=\frac{\bs{R}_{mk}^{\text{RF}}}{K_{mk}+1} - \hat{\bs{A}}_{mk}
\end{split}
\label{eq:tildeAmk}
\end{equation}
is the covariance matrix of $\tilde{\bs{g}}_{mk}$.

\subsection{Downlink payload data transmission}

Let us define $\bs{s}_d=\left[{s_d}_1 \ldots {s_d}_K\right]^T$ as the $K \times 1$ vector of symbols jointly transmitted to the $K$ \glspl{MS}, such that $E\left\{\bs{s}_d\bs{s}_d^H\right\}=\bs{I}_K$. Assuming the use of a centralized baseband precoder at the \gls{CPU}, symbol vector $\bs{s}_d$ undergoes some signal processing operations before being transmitted, including a power allocation process and a baseband precoding task at the \gls{CPU}, and an \gls{RF} precoding process at the \glspl{AP}. Thus, the transmitted signal vector from the $m$th active \gls{AP} can be generically expressed as
\begin{equation}
   \bs{x}_m=\bs{W}_m^{\text{RF}}\bs{W}_{d\,m}^{\text{BB}} \bs{\Upsilon}^{1/2} \bs{s}_d,
\label{eq:bsxm}
\end{equation}
with $\bs{W}_{d\,m}^{\text{BB}}=\left[\bs{w}_{dm1}^{\text{BB}}\ \ldots\ \bs{w}_{dmK}^{\text{BB}}\right]\ \in\ \mathbb{C}^{L_A \times K}$ denoting the baseband precoding matrix affecting the signal transmitted by the $m$th active \gls{AP}, and $\bs{\Upsilon}=\diag\left([\upsilon_1\,\ldots\,\upsilon_K]\right)$ being a $K \times K$ diagonal matrix containing the power control coefficients in its main diagonal. The power control coefficients are chosen to satisfy the power constraints
\begin{equation}
   \mathbb{E}\left\{\left\|\bs{x}_m\right\|^2\right\} = \sum_{k=1}^K \upsilon_k \theta_{mk}^{\text{BB/RF}} \leq \overline{P}_m,
\label{eq:Power_constraint}
\end{equation}
for all $m\in\mathcal{M}^A$, where we have used the definition
\begin{equation}
   \theta_{mk}^{\text{BB/RF}}=\mathbb{E}\left\{\left\|\bs{W}_m^{\text{RF}} \bs{w}_{dmk}^{\text{BB}}\right\|^2\right\},
\end{equation}
and $\overline{P}_m$ is the maximum average transmit power available at \gls{AP} $m$. Using this notation, the signal received by \gls{MS} $k$ can be expressed as
\begin{equation}
   {y_d}_k=\sum_{m\in\mathcal{M}^A} \breve{\bs{h}}_{mk}^T \bs{x}_m + {n_d}_k,
\label{eq:yk}
\end{equation}
where ${n_d}_k \sim \mathcal{CN}(0,\sigma_d^2)$ is the noise sample at \gls{MS} $k$.

The vector $\bs{y}_d = \left[{y_d}_1\, \ldots\, {y_d}_K\right]^T$ containing the signals received by the $K$ scheduled \glspl{MS} in the network can be written as
\begin{equation}
\begin{split}
   \bs{y}_d=&\sum_{m\in\mathcal{M}^A} \breve{\bs{H}}_m^T \bs{x}_m + \bs{n}_d \\
           =&\sum_{m\in\mathcal{M}^A} \breve{\bs{H}}_m^T \bs{W}_m^{\text{RF}} \bs{W}_{d\,m}^{\text{BB}} \bs{\Upsilon}^{1/2} \bs{s}_d  + \bs{n}_d \\
           =&\ \bs{G}^T \bs{W}_d^{\text{BB}}\bs{\Upsilon}^{1/2} \bs{s}_d + \bs{n}_d,
\end{split}
\label{eq:yd}
\end{equation}
where $\breve{\bs{H}}_m=\left[\breve{\bs{h}}_{m1}\, \ldots\, \breve{\bs{h}}_{mK}\right]\ \in\ \mathbb{C}^{L \times K}$ represents the \gls{MIMO} channel between the $m$th active \gls{AP} and the $K$ \glspl{MS}, $\bs{W}_d^{\text{BB}} = \left[{\bs{W}_{d\,m_1^A}^{\text{BB}}}^T\ \ldots\ {\bs{W}_{d\,m_{M_A}^A}^{\text{BB}}}^T\right]^T\ \in\ \mathbb{C}^{M_A L_A \times K}$ is the digital precoding filter stage implemented at the \gls{CPU}, $\bs{G}=[\bs{G}_{m_1^A}^T\,\ldots\,\bs{G}_{m_{M_A}^A}^T]^T\ \in\ \mathbb{C}^{M_A L_A \times K}$ is the global equivalent \gls{MIMO} channel (including the RF precoding/decoding matrices) between the $K$ \glspl{MS} and the digital processing stage at the \gls{CPU}, with $\bs{G}_m={\bs{W}_m^{\text{RF}}}^T \breve{\bs{H}}_m$ representing the equivalent \gls{MIMO} channel matrix between the $K$ \glspl{MS} and the digital processing stage corresponding to the $m$th active \gls{AP} and, finally, $\bs{n}_d=\left[{n_d}_1\, \ldots\, {n_d}_K\right]^T$ is the vector containing the noise samples at the \glspl{MS}.

Assuming the use of the classical \gls{ZF} \gls{MU-MIMO} baseband precoder to tackle the spatial multiplexing, we have that
\begin{equation}
   \bs{W}_d^{\text{BB}}=\hat{\bs{G}}^*\left(\hat{\bs{G}}^T \hat{\bs{G}}^*\right)^{-1}
\end{equation}
or, equivalently,
\begin{equation}
   \bs{W}_{d\,m}^{\text{BB}} = \hat{\bs{G}}_m^*\left(\hat{\bs{G}}^T \hat{\bs{G}}^*\right)^{-1}\ \forall m \in \mathcal{M}^A,
\end{equation}
where we have assumed that $\bs{G}=\hat{\bs{G}}+\tilde{\bs{G}}$ and $\bs{G}_m=\hat{\bs{G}}_m+\tilde{\bs{G}}_m$. Consequently, the signal received by the $k$th \gls{MS} can be expressed as
\begin{equation}
\begin{split}
   {y_d}_k = &\bs{g}_k^T \hat{\bs{G}}^*\left(\hat{\bs{G}}^T \hat{\bs{G}}^*\right)^{-1} \bs{\Upsilon}^{1/2}\bs{s}_d + {n_d}_k \\
           = &\left(\hat{\bs{g}}_k^T + \tilde{\bs{g}}_k^T\right) \hat{\bs{G}}^*\left(\hat{\bs{G}}^T \hat{\bs{G}}^*\right)^{-1} \bs{\Upsilon}^{1/2}\bs{s}_d + {n_d}_k \\
           = &\sqrt{\upsilon_k} {s_d}_k + \tilde{\bs{g}}_k^T \hat{\bs{G}}^*\left(\hat{\bs{G}}^T \hat{\bs{G}}^*\right)^{-1} \bs{\Upsilon}^{1/2}\bs{s}_d + {n_d}_k
\end{split}
\label{eq:ydk}
\end{equation}
The first term denotes the useful received signal, the second term contains the interference terms due to the use of imperfect \gls{CSI} (pilot contamination), and the third term is the thermal noise sample.

\subsection{Uplink payload data transmission}

In the \gls{UL}, the vector of received signals at the output of the $L_A$ \gls{RF} chains (including the \gls{RF} phase shifters) of the $m$th active \gls{AP} is given by
\begin{equation}
\begin{split}
   {\bs{r}_u}_m=&\sqrt{P_u}\sum_{k'=1}^K \bs{g}_{mk'} \sqrt{\omega_{k'}} {s_u}_{k'} + {\bs{n}_u}_m \\
               =&\sqrt{P_u}\bs{G}_m \bs{\Omega}^{1/2} \bs{s}_u + {\bs{n}_u}_m,
\end{split}
\end{equation}
where $P_u$ is the maximum average \gls{UL} transmit power available at any of the active \glspl{MS}, $\bs{s}_u=[{s_u}_1\,\ldots\,{s_u}_K]^T$ denotes the vector of symbols transmitted by the $K$ \gls{MS}, $\bs{\Omega}=\diag([\omega_1\,\ldots\,\omega_K])$, with $0 \leq \omega_k \leq 1$, is a matrix containing the power control coefficients used at the \glspl{MS}, and ${\bs{n}_u}_m \sim \mathcal{CN}(\bs{0},\sigma_u^2(N) \bs{I}_{L_A})$ is the vector of additive thermal noise samples at the output of the $L_A$ \gls{RF} chains of the $m$th active \gls{AP}. The received vector of signals at each of the active \glspl{AP} in $\mathcal{M}^A$ is forwarded to the \gls{CPU} where it is processed using a baseband combining matrix. In particular, assuming the use of \gls{ZF} \gls{MIMO} detection, the \gls{CPU} uses the detection matrix
\begin{equation}
   \bs{W}_u^{\text{BB}}=\left(\hat{\bs{G}}^H \hat{\bs{G}}\right)^{-1}\hat{\bs{G}}^H = {\bs{W}_d^{\text{BB}}}^T
\end{equation}
or, equivalently
\begin{equation}
   \bs{W}_{u m}^{\text{BB}}=\left(\hat{\bs{G}}^H \hat{\bs{G}}\right)^{-1}\hat{\bs{G}}_m^H = {\bs{W}_{d m}^{\text{BB}}}^T,\ \forall m \in \mathcal{M}^A,
\end{equation}
to jointly process the vector $\bs{z}_u=\left[{\bs{z}_u}_1^T\,\ldots\,{\bs{z}_u}_M^T\right]^T$ and obtain the vector of detected samples
\begin{equation}
\begin{split}
   \bs{y}_u=&\bs{W}_u^{\text{BB}}\bs{z}_u=\sqrt{P_u}\bs{W}_u^{\text{BB}} \bs{G} \bs{\Omega}^{1/2} \bs{s}_u + \bs{\eta}_u \\
           =&\sqrt{P_u} \bs{\Omega}^{1/2} \bs{s}_u + \sqrt{P_u}\bs{W}_u^{\text{BB}} \tilde{\bs{G}} \bs{\Omega}^{1/2} \bs{s}_u + \bs{\eta}_u,
\end{split}
\end{equation}
where $\bs{\eta}_u=\bs{W}_u^{\text{BB}}\bs{n}_u$. Again, the first term denotes the useful received signal, the second term contains the interference terms due to the use of imperfect \gls{CSI}, and the third term includes the thermal noise samples. The detected sample corresponding to the symbol transmitted by the $k$th \gls{MS} can then be obtained as
\begin{equation}
   {y_u}_k =\sqrt{P_u} \omega_k^{1/2} {s_u}_k + \sqrt{P_u}\left[\bs{W}_u^{\text{BB}} \tilde{\bs{G}} \bs{\Omega}^{1/2} \bs{s}_u\right]_k + {\eta_u}_k,
\label{eq:yuk}
\end{equation}
where $[\bs{x}]_k$ denotes the $k$th entry of vector $\bs{x}$.

\section{Performance metrics}
\label{sec:Performance_metrics}

\subsection{Spectral efficiency}
\label{sec:Achievable_rates}

Analysis techniques similar to those applied, for instance, in \cite{Hassibi03,Yang13,Interdonato16,Marzetta16,Ngo17,Nayebi17}, are used in this section to derive \gls{DL} and \gls{UL} spectral efficiencies (also known as achievable rates). In particular, the sum of the second and third terms on the \gls{RHS} of \eqref{eq:ydk}, for the \gls{DL} case, and \eqref{eq:yuk}, for the \gls{UL} case, are treated as \textit{effective noise}. The additive terms constituting the \textit{effective noise} are, in both \gls{DL} and \gls{UL} cases, mutually uncorrelated, and uncorrelated with ${s_d}_k$ and ${s_u}_k$, respectively. Therefore, both the desired signal and the so-called \textit{effective noise} are uncorrelated. Now, recalling the fact that uncorrelated Gaussian noise represents the worst case, from a capacity point of view, and that the complex-valued fast fading random variables characterizing the propagation channels between different pairs of \gls{AP}-\gls{MS} connections are independent, the \gls{DL} and \gls{UL} spectral efficiencies (measured in bits per second per Hertz) can be obtained as follows. The \gls{DL} spectral efficiency is given by
\begin{equation}
   {S_e}_d(\bs{\upsilon})=\sum_{k=1}^K {S_e}_{dk}(\bs{\upsilon})=\frac{\tau_d}{\tau_c}\sum_{k=1}^K \log_2\left(1+{\SINR_d}_k\right),
\end{equation}
with
\begin{equation}
   {\SINR_d}_k=\frac{\upsilon_k}{\sum_{k'=1}^K \upsilon_{k'} \varpi_{kk'} + \sigma_d^2},
   \label{eq:SINRdk}
\end{equation}
where
\begin{equation}
   \varpi_{kk'}=\left[\diag\left(\mathbb{E}\left\{{\bs{W}_d^{\text{BB}}}^H \tilde{\bs{g}}_k^* \tilde{\bs{g}}_k^T\ \bs{W}_d^{\text{BB}}\right\}\right)\right]_{k'}.
\end{equation}
Analogously, the \gls{UL} spectral efficiency is given by
\begin{equation}
   {S_e}_u(\bs{\omega})=\sum_{k=1}^K {S_e}_{uk}(\bs{\omega})=\frac{\tau_u}{\tau_c}\sum_{k=1}^K \log_2\left(1+{\SINR_u}_k\right),
\end{equation}
with
\begin{equation}
   {\SINR_u}_k=\frac{P_u \omega_k}{P_u \sum_{k'=1}^K \omega_{k'} \delta_{kk'} + \sigma_{\eta_{uk}}^2(N)},
   \label{eq:SINRuk}
\end{equation}
where
\begin{equation}
   \delta_{kk'}=\left[\diag\left(\mathbb{E}\left\{\bs{W}_u^{\text{BB}} \tilde{\bs{g}}_{k'} \tilde{\bs{g}}_{k'}^H {\bs{W}_u^{\text{BB}}}^H\right\}\right)\right]_k,
\end{equation}
and
\begin{equation}
   \sigma_{\eta_{uk}}^2(N)=\sigma_u^2(N) \left[\diag\left(\mathbb{E}\left\{\bs{W}_u^{\text{BB}} {\bs{W}_u^{\text{BB}}}^H\right\}\right)\right]_k.
\end{equation}

\subsection{Power consumption model}
\label{sec:power_consumption}

In a cell-free massive \gls{MIMO} network implementing an \gls{ASO} strategy, \glspl{AP} can be either in \textit{active} or \textit{sleep} mode. Moreover, an \gls{AP} in active mode can be either receiving signals during the \gls{UL} training and payload data transmission phases, or transmitting information during the \gls{DL} payload data transmission phase. When in active mode, the power consumption at the $m$th \gls{AP} depends on the \gls{UL} spectral efficiency ${S_e}_u(\bs{\omega})$ during the \gls{UL} payload data transmission phase or on the radiated power $P_m^{\text{tx}}$ during the \gls{DL} payload data transmission phase. But it also depends on parameters such as the efficiency of the power amplifier, the power consumed by the small-signal \gls{RF} transceiver and the baseband circuitry, or the losses produced by the feeder, the DC-DC power supply, the main supply, or the cooling system \cite{Auer11,Tombaz11,Desset12,Bjornson16c,Dai16}. When in the sleep mode, the \gls{AP} is kept in a state allowing a fairly rapid activation and hence it is not completely turned off. It is thus in a reduced power consumption state in which, although it is not radiating or receiving power, there are components such as the power supply, some of the signal processing blocks, and part of the cooling system that are still active and thus consuming power. Accordingly, a linear model can be used to approximate the total power consumed at the $m$th \gls{AP} as (see, for instance, \cite{Auer11,Tombaz11,Desset12,Bjornson16c,Dai16} and references therein)
\begin{equation}
   P_m^{\text{AP}}=\begin{cases}
                      \frac{P_m^{\text{tx}}(\bs{\upsilon})}{\alpha_m^{\text{AP}}} + P_{m\,d}^{\text{AP,fix}} + L_A P_{m\,d}^{\text{AP,chain}} & \text{DL Active} \\
                      B \xi_m^{\text{AP}} {S_e}_u(\bs{\omega}) + P_{m\,u}^{\text{AP,fix}} + L_A P_{m\,u}^{\text{AP,chain}} & \text{UL Active} \\
                      P_{m\,\text{sleep}}^{\text{AP,fix}} + L_A P_{m\,\text{sleep}}^{\text{AP,chain}} & \text{Sleep},
                   \end{cases}
\end{equation}
where $\alpha_m^{\text{AP}}$ is the power amplifier efficiency at the $m$th \gls{AP}, $B$ is the system bandwidth, $\xi_m^{\text{AP}}$ is the traffic-dependent power consumption coefficient (in Watt per bit/s), $P_{m\,d}^{\text{AP,fix}}$ and $P_{m\,u}^{\text{AP,fix}}$ denote, respectively, the \gls{DL} and \gls{UL} power consumption figures that are independent of both the number of \gls{RF} chains and the traffic load, $P_{m\,d}^{\text{AP,chain}}$ and $P_{m\,u}^{\text{AP,chain}}$ model the \gls{DL} and \gls{UL} traffic-independent power consumed by the circuitry related to each \gls{RF} chain of the $m$th \gls{AP}, respectively and, finally, $P_{m\,\text{sleep}}^{\text{AP,fix}}$ and $P_{m\,\text{sleep}}^{\text{AP,chain}}$ are the \gls{RF} chain-independent and \gls{RF} chain-dependent power consumed by the $m$th \gls{AP} when in sleep mode.

A similar power consumption model can be established for the fronthaul links connecting the \glspl{AP} to the \gls{CPU}. In particular, the power consumed by the $m$th fronthaul link when in active mode depends on the amount of traffic it has to convey and, thus, the total power consumption can be approximated as \cite{Nguyen17,Ngo18}
\begin{equation}
   P_m^{\text{FH}}=\begin{cases}
                      B \xi_m^{\text{FH}} {S_e}_d(\bs{\upsilon}) + P_m^{\text{FH,fix}} & \text{DL Active} \\
                      B \xi_m^{\text{FH}} {S_e}_u(\bs{\omega}) + P_m^{\text{FH,fix}} & \text{UL Active} \\
                      P_{m\,\text{sleep}}^{\text{FH,fix}} & \text{Sleep},
                   \end{cases}
\end{equation}
where $\xi_m^{\text{FH}}$ is the traffic-dependent power consumption coefficient (in Watt per bit/s), $P_m^{\text{FH,fix}}$ is the traffic-independent power consumption when in active mode, and $P_{m\,\text{sleep}}^{\text{FH,fix}}$ accounts for the power consumed by the $m$th fronthaul link when in sleep mode.

The power consumption model for the \glspl{MS} can also be approximated as
\begin{equation}
   P_k^{\text{MS}}=\begin{cases}
                      B\xi_k^{\text{MS}} {S_e}_{dk}(\bs{\upsilon}) + P_{k\,d}^{\text{MS,fix}} & \text{DL} \\
                      \frac{P_u \omega_k}{\alpha_k^{\text{MS}}} + P_{k\,u}^{\text{MS,fix}} & \text{UL},
                   \end{cases}
\end{equation}
where, again, $\alpha_k^{\text{MS}}$ is the power amplifier efficiency at the $k$th \gls{MS}, $\xi_k^{\text{MS}}$ is the traffic-dependent power consumption coefficient (in Watt per bit/s), $P_{k\,d}^{\text{MS,fix}}$ and $P_{k\,u}^{\text{MS,fix}}$ model the power consumed by the internal circuitry of the \gls{MS} independently of the average radiated power, and ${S_e}_{dk}(\bs{\upsilon})$ denotes the \gls{DL} spectral efficiency of the $k$th \gls{MS}.

Putting all the pieces together, the total power consumption of the cell-free massive-\gls{MIMO} network can be modeled as
\begin{equation}
\begin{split}
   {P_T}_d(\bs{\upsilon})&=P_{T d}^{\text{fix}}+B\sum_{k=1}^K \xi_k^{\text{MS}} {S_e}_{dk}(\bs{\upsilon})\\
                         &+\sum_{m=1}^M \left(\frac{\tau_d}{\tau_c}\frac{P_m^{\text{tx}}(\bs{\upsilon})}{\alpha_m^{\text{AP}}} + B \xi_m^{\text{FH}} {S_e}_d(\bs{\upsilon})\right),
\end{split}
\end{equation}
for the \gls{DL} payload data transmission phase, and as
\begin{equation}
\begin{split}
   {P_T}_u(\bs{\omega})&=P_{T u}^{\text{fix}}+\sum_{k=1}^K \frac{\tau_u}{\tau_c}\frac{P_u \omega_k}{\alpha_k^{\text{MS}}}\\
                       &+B \sum_{m=1}^M \left(\xi_m^{\text{AP}} + \xi_m^{\text{FH}}\right) {S_e}_u(\bs{\omega}),
\end{split}
\end{equation}
for the \gls{UL} payload data transmission phase, with
\begin{equation}
\begin{split}
    P_{T l}^{\text{fix}}&=\frac{\tau_l}{\tau_c}\left[\sum_{k=1}^K P_{k\,l}^{\text{MS,fix}}\right. \\
    &+\sum_{m=1}^M \left(P_m^{\text{FH,fix}} + P_{m\,l}^{\text{AP,fix}} + L_A P_{m\,l}^{\text{AP,chain}}\right) \\
    &+\left.\sum_{m=1}^M \left(P_{m\,\text{sleep}}^{\text{FH,fix}} + P_{m\,\text{sleep}}^{\text{AP,fix}} + L_A P_{m\,\text{sleep}}^{\text{AP,chain}}\right)\right]
\end{split}
\end{equation}
where $l$ has been used as a token to represent either the \gls{DL} ($l=d$) or the \gls{UL} ($l=u$). As stated by Desset \textit{et al.} in \cite{Desset12}, although this simple linear model is not designed to provide very accurate absolute figures, it will enable a fair comparison among different on/off switching strategies for green cell-free massive-\gls{MIMO} networking.

\subsection{Energy efficiency}
\label{sec:energy_efficiency}

The energy efficiency during the \gls{DL} and \gls{UL} payload data transmission phases can be expressed as
\begin{equation}
   {E_e}_d(\bs{\upsilon})=\frac{B {S_e}_d(\bs{\upsilon})}{{P_T}_d(\bs{\upsilon})}
\end{equation}
and
\begin{equation}
   {E_e}_u(\bs{\omega})=\frac{B {S_e}_u(\bs{\omega})}{{P_T}_u(\bs{\omega})},
\end{equation}
respectively. We can also define a \textit{weighted} energy efficiency metric as
\begin{equation}
   {E_e}(\bs{\upsilon},\bs{\omega})=(1-\mu) {E_e}_d(\bs{\upsilon}) + \mu {E_e}_u(\bs{\omega}),
\end{equation}
where $0\leq \mu \leq 1$ is a weighting coefficient allowing for the control of a trade-off between \gls{DL} and \gls{UL} energy efficiencies.

\section{\gls{AP} switching strategies based on goodness-of-fit}
\label{sec:ASO_strategies}

Optimal \gls{ASO} strategies aim at activating the subset of $M_A$ \glspl{AP} providing the maximum energy efficiency. Determining the optimal subset of \glspl{AP}, however, is an NP-hard problem that calls for the evaluation of the performance provided by all possible combinations of $M_A$ out of $M$ \glspl{AP}. Hence, the implementation of computationally feasible selection strategies will only be possible by relying on the development of heuristic suboptimal algorithms. In the following we describe some heuristic \gls{ASO} strategies that are based on the \gls{GoF} theory. Under ideal conditions, the set of selected \glspl{AP} should be adapted to scenario variations due to, among others, changes in the number and/or location of \glspl{MS} or changes in the geographical distribution of shadow fading. In most practical situations, however, these variations occur too quickly so as to allow the implementation of realistic \gls{ASO} schemes that can adapt to them. The \gls{GoF}-based \gls{ASO} strategies have been specifically designed to cope with long-term non-uniform spatial traffic densities. In particular, it seems intuitively satisfactory to try to match the spatial distribution of active \glspl{AP} to that of the \glspl{MS} in the network in an attempt to selectively activate that parts of the network where most likely active users are located. This is the rational behind the \gls{GoF} methods presented next. The performance provided by these \gls{ASO} schemes will be benchmarked against that provided by three of the \gls{ASO} strategies previously proposed in \cite{femenias2020access}, that have been suitably adapted to the \gls{mmWave} scenario.
:
\begin{itemize}
\item \textit{\Gls{RS-ASO}}: Active \glspl{AP} are randomly selected and the only parameter that is optimized in trying to maximize the energy efficiency of the network is the number of active \glspl{AP} as a function of the number (or spatial density) of \glspl{MS} in the serviced area. The energy efficiency performance improvement provided by the elementary \gls{RS-ASO} strategy will serve as a lower bound on the performance improvement any other sensible \gls{ASO} scheme may bring along.

\item \textit{\Gls{MPL-ASO}}: The set $\mathcal{M}^A$ of active \glspl{AP} is selected based on large-scale propagation losses between \glspl{AP} and \glspl{MS}. For those cases in which $M_A \geq K$, the algorithm selects (in an ordered manner) the group of $M_A$ \glspl{AP} showing the minimum propagation losses to the $K$ \glspl{MS} in the network. For those cases in which $M_A < K$, instead, a set of $M_A$ \textit{virtual} \glspl{MS} is first generated by relying on the k-means clustering method and the previously described procedure is then applied to the virtual \glspl{MS} to select the set of active \glspl{AP}. A detailed explanation of this \gls{ASO} strategy can be found in \cite{femenias2020access}. It is important to note at this point that the pace at which the \glspl{AP} would have to be switched on/off under this strategy would be so high that it would be hardly implementable in practice.

\item \textit{\Gls{OG-ASO}}: This is an iterative greedy algorithm that, starting with the $M$ available \glspl{AP} in the first iteration, in the $i$th iteration of the algorithm evaluates the $(M+1-i)$ possible configurations of $(M-i)$ active \glspl{AP} resulting from switching off one of them, and selects the configuration maximizing the energy efficiency. The algorithm iterates until obtaining the configuration of active \glspl{AP} maximizing the energy efficiency of the network. The energy efficiency performance improvement provided by this \textit{unrealistic} \gls{ASO} strategy will serve as an upper bound against which to compare the performance provided by the other \gls{ASO} schemes.
\end{itemize}

\subsection{Motivation for goodness-of-fit}

When a particular probability distribution has been specified to model a random phenomenon (such as the spatial distribution of \glspl{MS}) the validity of the specified or assumed distribution model may be statistically verified or disproved by using \gls{GoF} tests \cite{dAgostino86,aslan2002comparison,evans2008distribution,Williams10,read2012goodness}. In this context, we can use \gls{GoF} techniques to determine which \glspl{AP} should be turned on or off in such a way that the resulting \gls{AP} distribution matches the non-uniform \gls{MS} distribution.

As previously described in Section \ref{subsec:Spatial_modeling}, we consider that the target region is tessellated in a regular grid of $N_X$ by $N_Y$ rectangular pixels, and the probability density of \glspl{MS} on pixel $(x,y)$ is denoted as $f_{x,y}^{MS}$. For a given set $\mathcal{M}^A$ of active \glspl{AP}, we can determine an \textit{estimate} of the probability density of \glspl{AP} $f_{x,y}^{AP}$ on this particular pixel as
\begin{equation}
   f_{x,y}^{\gls{AP}}=\frac{M_A^{(x,y)}}{M_A},
   \label{eq:APpdf}
\end{equation}
where $M_A^{(x,y)}$ is the number of active \glspl{AP} on pixel $(x,y)$, and $M_A$ is the number of active \glspl{AP} in the target region. Thus, relying on \eqref{eq:APpdf}, the \gls{GoF} techniques can establish a link between the spatial distribution of \glspl{MS} (i.e., $f_{x,y}^{\gls{MS}}$ values) and the spatial distribution of active \glspl{AP} (i.e., $f_{x,y}^{\gls{AP}}$ values). In particular, in the following subsections three novel \gls{ASO} strategies are proposed: two of them are based on widely applied \gls{GoF} techniques \cite{dAgostino86}, namely, the Chi-square test, and the Kolmogorov-Smirnov test, and the third one is based on the concept of \textit{statistical energy}, described by Aslan and Zech in \cite{Aslan05}. The optimal number of active \glspl{AP} (under any of the proposed \gls{GoF}-based \gls{ASO} strategies) when serving a given amount of \glspl{MS} would be the one providing the maximum energy efficiency.

\subsection{Chi-square based \gls{ASO}}
The \gls{ChiS} is closely connected to a least square fit between the observed normalized frequencies of \glspl{AP} per pixel $\{f_{x,y}^{\gls{AP}}\}_{\forall (x,y)}$ with the corresponding theoretical probability densities $\{f_{x,y}^{\gls{MS}}\}_{\forall (x,y)}$. Given a set $\mathcal{M}^A$ of selected active \glspl{AP}, the \gls{GoF} metric implemented by the \gls{ChiS} test can be expressed as
 \begin{equation}
   D_{\mathcal{M}^A}^{(\gls{ChiS})} = \sum_{x=1}^{N_X} \sum_{y=1}^{N_Y} \frac{\left( f_{x,y}^{\gls{AP}} - f_{x,y}^{\gls{MS}} \right)^2}{f_{x,y}^{\gls{MS}}}.
   \label{eq:ChiS_ASO}
\end{equation}
The lower the value of $D_{\mathcal{M}^A}^{(\gls{ChiS})}$ the better the \gls{GoF} between the spatial distributions of \glspl{MS} and active \glspl{AP}. Hence, by using \eqref{eq:ChiS_ASO} the optimal \gls{ChiS-ASO} algorithm would be the one selecting the \glspl{AP} whose corresponding $D_{\mathcal{M}^A}^{(\gls{ChiS})}$ value is minimum. Having a large number of \glspl{AP} in the network, however, NP-hardness forbids the implementation of a brute force algorithm to solve this optimization problem. Consequently, an iterative \gls{ChiS-ASO} algorithm is proposed that, starting with a set containing all the \glspl{AP} in the network, in each iteration switches-off the single \gls{AP} leading to the minimum $D_{\mathcal{M}^A}^{(\gls{ChiS})}$ value when removed.

\subsection{Kolmogorov-Smirnov based \gls{ASO}}

The \gls{ChiS} test just described is an important special case of the power-divergence statistic \cite{read2012goodness} and is computationally simple. Unfortunately, however, it also has a serious drawback: as it neglects the correlation between adjacent elements of the histogram (i.e., between adjacent pixels), it exhibits a rather poor performance in detecting slowly varying deviations between both the analytical and predicted statistical distributions \cite{aslan2002comparison}. Furthermore, the \gls{ChiS} scheme requires of a large number of intervals and/or samples, that is, it requires of large $M_A$, $N_X$ and $N_Y$ values. In this regard, the \gls{KS}-based approach has some advantages over the \gls{ChiS} test. In particular, with the \gls{KS}-based strategy the problem associated with small number of intervals and/or samples would not be an issue \cite{aslan2002comparison}. Moreover, the \gls{KS-ASO} algorithm takes into consideration the possible correlations between adjacent elements (or pixels) by using the \glspl{CDF} in calculating the discrepancy metric.

For the one-dimensional case, the \gls{KS} test considers simply the largest absolute difference between the two \glspl{CDF} as a measure of misfit. In this case, the result of the test is independent of the direction of ordering of the data, that is, it is independent of whether we consider the cumulative probabilities $P(x>X)$ or $P(x<X)$. In a multi-dimensional case, however, defining the \gls{CDF} as $P(x<X, y<Y, \ldots)$ is ambiguous since the directions in which we choose to order the different random variables are arbitrary. In fact, in an $n$-dimensional case there are $2^n-1$ independent ways of defining the \gls{CDF}. A straightforward way to avoid the dependency of the \gls{KS} test on the particular orderings chosen is to specify the discrepancy metric as the largest absolute difference between empirical and theoretical \glspl{CDF} when all possible ordering combinations (i.e., $2^n$) are considered \cite{Peacock83,Fasano87}. In our particular two-dimensional case, this corresponds to recognizing that the statistical descriptions of the spatial location of both \glspl{AP} and \glspl{MS} in all four quadrants of the plane defined by $(x< X, y < Y)$, $(x < X, y > Y)$, $(x> X, y < Y)$ and $(x> X, y > Y)$ are equally valid, and that the discrepancy metric can be obtained as the largest of the four differences in empirical and theoretical \glspl{CDF}. This can be mathematically expressed as
\begin{equation}
   D_{\mathcal{M}^A}^{(\gls{KS})}= \max_{i\in\{1,2,3,4\}} \left\{\max_{x,y} \left|F_{x,y}^{\gls{AP},i} - F_{x,y}^{\gls{MS},i}\right|\right\},
   \label{eq:KS_ASO}
\end{equation}
where the theoretical and empirical \glspl{CDF} describing the spatial distributions of \glspl{MS} and \glspl{AP}, respectively, in the four quadrants of the plane can be obtained from $f_{x,y}^{\gls{MS}}$ in \eqref{eq:MS_pdf} and $f_{x,y}^{\gls{AP}}$ in \eqref{eq:APpdf} as
\begin{equation}
\begin{split}
   &F_{x,y}^{CE,1}=\sum_{i=1}^x \sum_{j=1}^y f_{i,j}^{CE},\qquad F_{x,y}^{CE,2}=\sum_{i=1}^x \sum_{j=y}^{N_Y} f_{i,j}^{CE}, \\
   &F_{x,y}^{CE,3}=\sum_{i=x}^{N_X} \sum_{j=1}^y f_{i,j}^{CE},\qquad F_{x,y}^{CE,4}=\sum_{i=x}^{N_X} \sum_{j=y}^{N_Y} f_{i,j}^{CE},
\end{split}
\end{equation}
with $CE$ denoting a token used to represent one of the communication ends, either the \gls{AP} (i.e., $CE=\gls{AP}$) or the \gls{MS} (i.e., $CE=\gls{MS}$).

Using, once again, a greedy strategy starting with a set containing all the \glspl{AP} in the network, an iterative \gls{KS-ASO} algorithm switches-off, in each iteration, the single \gls{AP} resulting in the minimum $D_{\mathcal{M}^A}^{(\gls{KS})}$ value when removed.

\subsection{Statistical energy-based \gls{ASO}}

In \cite{Aslan05}, Aslan and Zech introduce the concept of \textit{statistical energy} as a tool for multivariate \gls{GoF} tests. Similar to what is done when dealing with electric charge distributions, where charges of opposite sign are in a state of minimum energy when they are equally distributed, they define the \textit{statistical energy} of statistical distributions and use it to mathematically describe a \gls{GoF} test.

For a given set $\mathcal{M}^A=\left\{m_1^A,\ldots,m_{M_A}^A\right\}$ of active \glspl{AP}, the statistical energy-based test aims at comparing the sample of spatial locations of these \glspl{AP} (which follow an unknown \gls{pdf}) to the reference theoretical spatial \gls{pdf} $f_{x,y}^{\gls{MS}}$ of the \glspl{MS}. To this end, let us first define the location (on a complex plane) of \gls{AP} $m_l^A$, for all $l\in\{1,\ldots,M_A\}$, as $p_{m_l^A}^{\gls{AP}}$, with \textit{charge} $1/M_A$. Furthermore, let us use the locations $p_{x,y}$ of pixels $(x,y)$ for all $x\in\{1,\ldots,N_X\}$ and $y\in\{1,\ldots, N_Y\}$, which conform to the spatial \gls{pdf} $f_{x,y}^{\gls{MS}}$ in \eqref{eq:MS_pdf}. Using these components, the statistical energy test statistic in \cite[(3.1)]{Aslan05} can be adapted to our problem at hand as
\begin{equation}
\begin{split}
   &D_{\mathcal{M}^A}^{(LSE)}=\frac{1}{M_A(M_A-1)}\sum_{l=1}^{M_A}\sum_{n=l+1}^{M_A}R_{\log}\left(\left|p_{m_l^A}^{AP}-p_{m_n^A}^{AP}\right|\right) \\
   &\quad-\frac{1}{M_A}\sum_{x=1}^{N_X}\sum_{y=1}^{N_Y} f_{x,y}^{MS} \sum_{l=1}^{M_A} R_{\log}\left(\left|p_{x,y}-p_{m_l^A}^{AP}\right|\right),
\end{split}
\end{equation}
where the logarithmic distance $R_{\log}(r)=-\ln(r+\varepsilon)$, with $\varepsilon=1/(2 N_X N_Y \max\{f_{x,y}^{MS}\})$, is used because, as shown in \cite[Fig.2]{Aslan05}, it provides a reasonably good performance in scenarios with very dissimilar spatial distributions (background contaminations in the notation used by Aslan and Zech in \cite{Aslan05}). Again, a greedy strategy is implemented that, starting with a set containing all the \glspl{AP} in the network, implements an iterative \gls{LSE-ASO} algorithm that switches-off, in each iteration, the single \gls{AP} resulting in the minimum $D_{\mathcal{M}^A}^{(LSE)}$ value when removed.

\section{Numerical results}
\label{sec:numerical_results}

This section presents a comprehensive set of numerical results to qualitatively and quantitatively assess the performance of the proposed \gls{ASO} strategies in a cell-free mmWave massive MIMO context in terms of both energy and spectral efficiencies, and also overall power consumption. Particular attention is paid to the effects caused by modifying the \gls{RF} infrastructure at the \glspl{AP} and the consequences of changing the density of \glspl{MS} per area unit and their corresponding spatial distribution. As generally done in most cell-free background literature, an scenario is considered where \glspl{AP} are initially uniformly distributed at random within a square coverage area of side $D$ whose boundaries are wrapped around, thus effectively simulating the operation of a network without boundary effects. Unlike most previous works, however, the positions of the \glspl{MS} follow a non-uniform distribution as previously explained in Section \ref{subsec:Spatial_modeling}.

The main parameters used throughout all simulations in this section are collected in Table \ref{tab:parameters}. These parameters have been borrowed from a variety of prior research works (see, for instance, \cite{3GPPmmWaveR14,Neil17,Nguyen17,Auer11,Bjornson16c,Femenias19,Bjornson19}). Results shown next have been obtained using the heuristic power allocation introduced by Nayebi \textit{et al.} in \cite[eq. (21)]{Nayebi17} (i.e., $\upsilon_k=P_d/\left(\max_m\sum_{k'=1}^K \theta_{mk'}\right)$ for all $k$) for the \gls{DL} case, which can be deemed as a computationally simple approximation to the max-min power allocation approach, and a full-power transmission strategy (i.e., $\omega_k=$1 for all $k$) for the \gls{UL} case. Nonetheless, it should be stressed that the proposed framework is indeed applicable to any other power allocation policy (such as those presented in, for example, \cite{Ngo17,Nayebi17,Ngo18}). Finally, and following the work in \cite{Femenias19}, a balanced random pilot assignment scheme is implemented whereby \glspl{MS} are allocated pilot sequences that are sequentially and cyclically selected from the ordered set of available orthogonal pilots.

\begin{table}[t!]
\renewcommand{\arraystretch}{1.1}
\caption{\small Summary of default simulation parameters}
\label{tab:parameters}
\centering
\begin{tabular}{l|c}
\hline

\bfseries Parameters & \bfseries Value\\

\hline

\footnotesize {Carrier frequency: $f_0$} & \footnotesize {28 GHz}\\
\footnotesize {Bandwidth: $B$} & \footnotesize {20 MHz}\\
\footnotesize {Side of the square coverage area: $D$} & \footnotesize {500 m}\\
\footnotesize {AP/MS antenna height: $h_{\text{AP}}/h_{\text{MS}}$} & \footnotesize {10/1.65 m}\\
\footnotesize {Noise figure at the MS: ${NF}_{\text{MS}}$} & \footnotesize {9 dB} \\
\footnotesize {Noise figure of the LNA at the AP: ${NF}_{\text{LNA}}$} & \footnotesize {1.6 dB} \\
\footnotesize {Gain of the LNA at the AP: $G_{\text{LNA}}$} & \footnotesize {22 dB} \\
\footnotesize {Phase splitters attenuation of the at the AP: $L_{\text{PS}}$} & \footnotesize {3 dB}\\
\footnotesize {Power combiner attenuation at the AP: $L_{\text{PC}_{in}}$} & \footnotesize {3 dB}\\
\footnotesize {Noise figure of the RF chain at the AP: ${NF}_{\text{RF}}$} & \footnotesize {7 dB}\\
\footnotesize {AP maximum transmit power: $P_d$} & \footnotesize {200 mW}\\
\footnotesize {MS maximum transmit power: $P_u=P_p$} & \footnotesize {100 mW}\\
\footnotesize {Antenna configuration at each AP: $N_x \times N_y$} & \footnotesize {8 $\times$ 1}\\
\footnotesize {Minimum separation between antenna elements} & \footnotesize {$\lambda/2$}\\
\footnotesize {Coherence interval length: $\tau_c$} & \footnotesize {200 samples}\\
\footnotesize {Training phase length: $\tau_p$} & \footnotesize {20 samples}\\
\footnotesize {Parameters for MS condition: $a_{\text{out}}$, $b_{\text{out}}$, $a_{\text{LOS}}$} & \footnotesize{1/30, 5.2, 1/67.1} \\
\footnotesize {Pathloss parameters Case LOS: $\alpha$, $\beta$, $\sigma_\chi$} & \footnotesize{61.34, 2.10, 4.0 dB} \\
\footnotesize {Pathloss parameters Case NLOS: $\alpha$, $\beta$, $\sigma_\chi$} & \footnotesize{61.34, 3.19, 8.2 dB} \\
\footnotesize {Shadow fading decorrelation distance: $d_{\text{dcorr}}$} & \footnotesize {9 m}\\
\footnotesize {Shadow fading correlation among \glspl{AP}:} & \footnotesize {0.5}\\
\footnotesize {Ricean $K$-factor distribution: $\mu_K$/$\sigma_K$} & \footnotesize {9/5 dB}\\
\footnotesize {Number of clusters LOS/NLOS: $C_{mk}$} & \footnotesize {12/19}\\
\footnotesize {Number of paths per cluster LOS/NLOS: $P_{mk}$} & \footnotesize {20/20}\\
\footnotesize {Azimuth angular spread (AP) LOS/NLOS: $\lambda_{\text{rms}}^{-1}$} & \footnotesize {3$^o$/10$^o$}\\
\footnotesize {Elevation angular spread: $\lambda_{\text{rms}}^{-1}$} & \footnotesize {7$^o$}\\
\footnotesize {Cluster power fraction parameters: $r_\tau/\zeta$} & \footnotesize {3/4}\\
\footnotesize {Power amplifier efficiency: $\alpha_m^{\text{AP}}/\alpha_k^{\text{MS}}$} & \footnotesize {0.39/0.3}\\
\footnotesize {Power per \gls{AP}/\gls{MS}/FH traffic: $\xi_m^{\text{AP}}/\xi_k^{\text{MS}}/\xi_m^{\text{FH}}$} & \footnotesize {0.25/0.25/0.25 $\frac{\text{W}}{\text{Gbps}}$}\\
\footnotesize {\gls{AP} fixed power: $P_{m\,d}^{\text{AP,fix}}/P_{m\,u}^{\text{AP,fix}}$} & \footnotesize {8/6 W}\\
\footnotesize {\gls{AP} fixed power per RF chain: $P_{m\,d}^{\text{AP,chain}}/P_{m\,u}^{\text{AP,chain}}$} & \footnotesize {0.2/0.15 W}\\
\footnotesize {\gls{AP} fixed power (sleep): $P_{m\,\text{sleep}}^{\text{AP,fix}}$} & \footnotesize {0.8 W}\\
\footnotesize {\gls{AP} fixed power per RF chain (sleep): $P_{m\,\text{sleep}}^{\text{AP,chain}}$} & \footnotesize {0.02 W}\\
\footnotesize {\gls{MS} fixed power: $P_{k\,d}^{\text{MS,fix}}/P_{k\,u}^{\text{MS,fix}}$} & \footnotesize {1/0.75 W}\\
\footnotesize {FH fixed power: $P_m^{\text{FH,fix}}$} & \footnotesize {5 W}\\
\footnotesize {FH fixed power (sleep): $P_{m\,\text{sleep}}^{\text{FH,fix}}$} & \footnotesize {0.5 W}\\
\hline
\end{tabular}
\end{table}

\begin{figure}[t!]
    \centering
    \includegraphics[width=\linewidth]{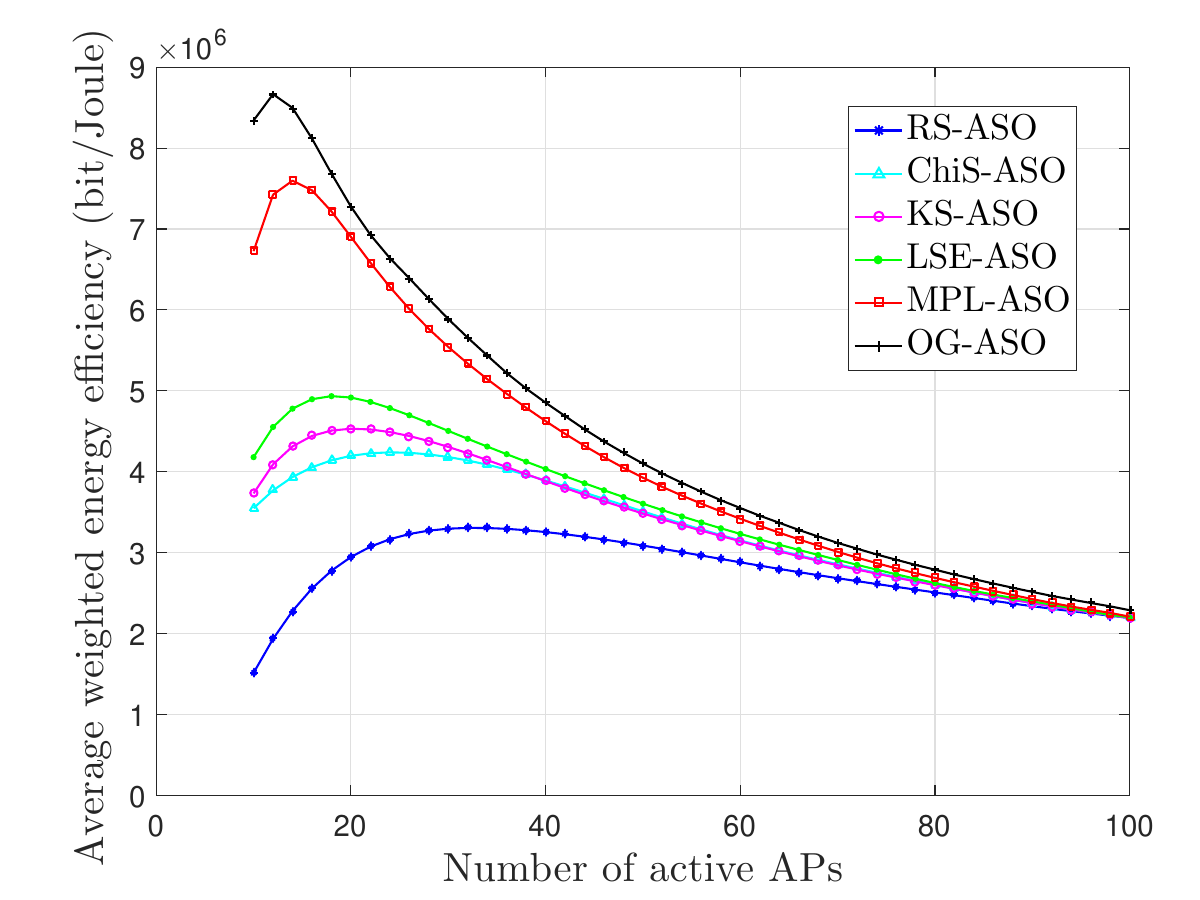}
    \caption{\small Impact of the \gls{ASO} strategy on the average (equally) weighted energy efficiency as a function of the number of active \glspl{AP} (weighted \gls{UL}/\gls{DL} scenario with $\mu=0.5$).}
    \label{fig:Weighted_ASO_strategy_EE}
\end{figure}

\begin{figure*}[t!]
    \centering
    \begin{subfigure}[t]{0.32\textwidth}
        \centering
        \includegraphics[height=6.8cm]{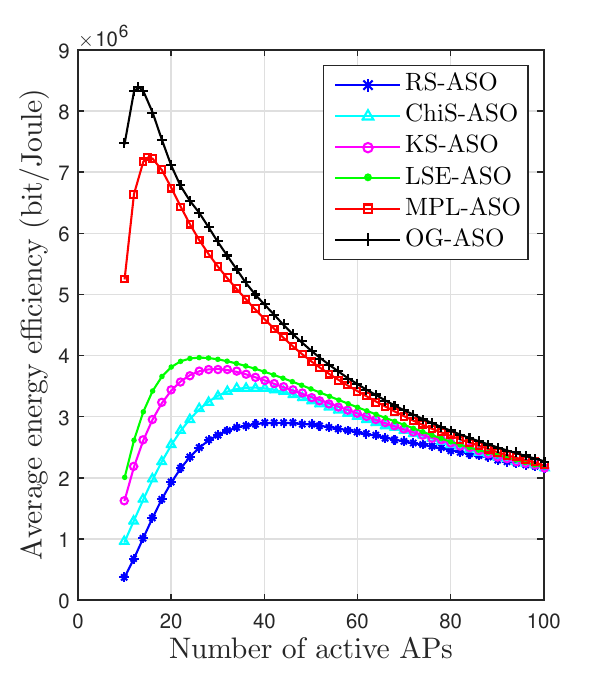}
        \label{fig:Fig_DL_ASO_strategy_EE}
    \end{subfigure}
    \begin{subfigure}[t]{0.32\textwidth}
        \centering
        \includegraphics[height=6.8cm]{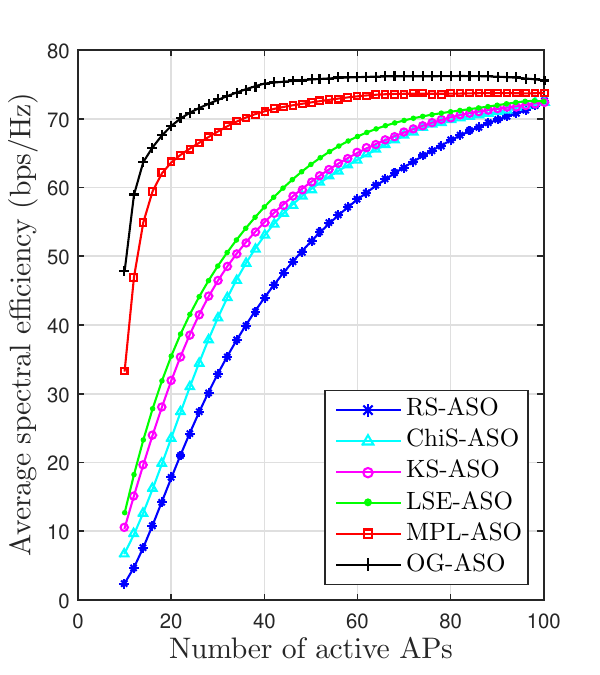}
        \label{fig:Fig_DL_ASO_strategy_ES}
    \end{subfigure}
    \begin{subfigure}[t]{0.32\textwidth}
        \centering
        \includegraphics[height=6.8cm]{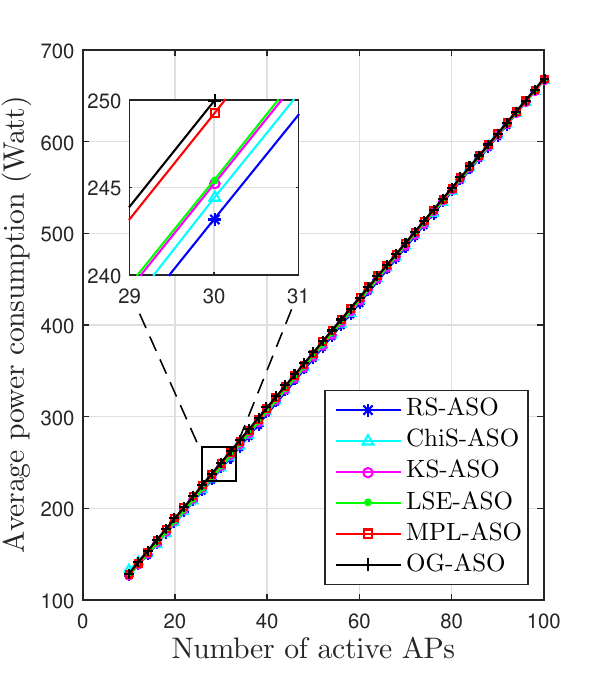}
        \label{fig:Fig_DL_ASO_strategy_PC}
    \end{subfigure}

    \caption{\small Impact of the \gls{ASO} strategy on the \gls{DL} average energy efficiency, spectral efficiency and power consumption as a function of the number of active \glspl{AP}.}
    \label{fig:var_ASO_strategy_DL}
\end{figure*}

\begin{figure*}[t!]

    \centering
    \begin{subfigure}[t]{0.32\textwidth}
        \centering
        \includegraphics[height=6.8cm]{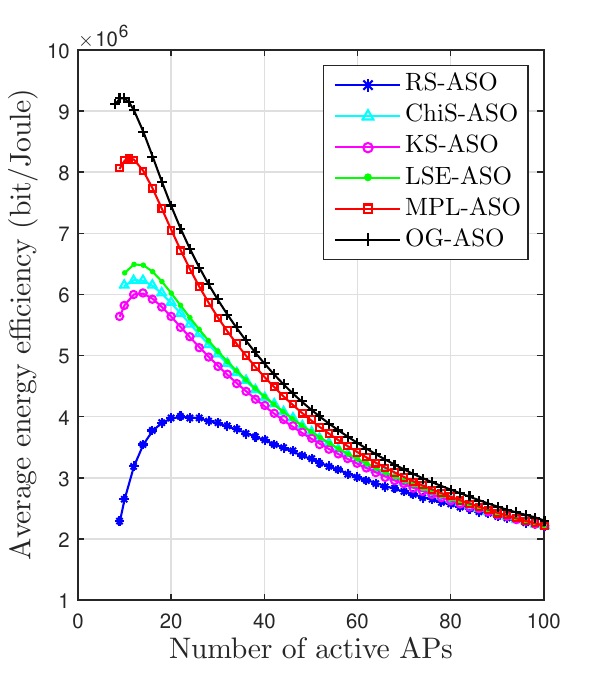}
        \label{fig:Fig_UL_ASO_strategy_EE}
    \end{subfigure}
    \begin{subfigure}[t]{0.32\textwidth}
        \centering
        \includegraphics[height=6.8cm]{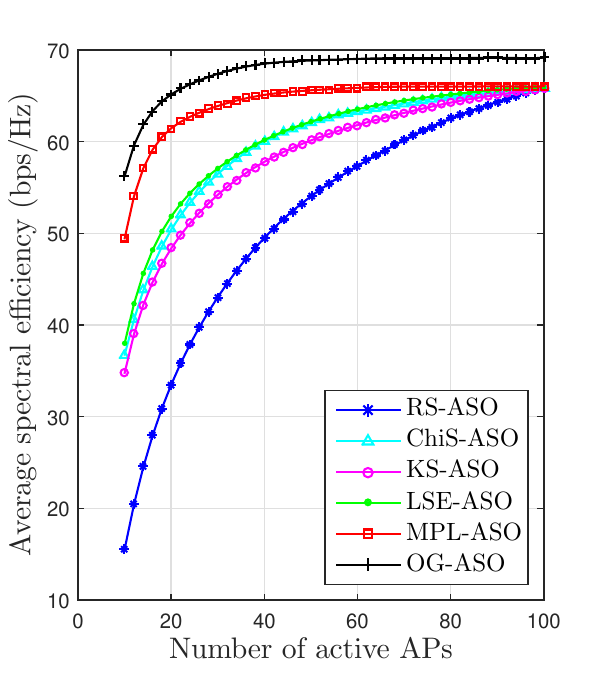}
        \label{fig:Fig_UL_ASO_strategy_ES}
    \end{subfigure}
    \begin{subfigure}[t]{0.32\textwidth}
        \centering
        \includegraphics[height=6.8cm]{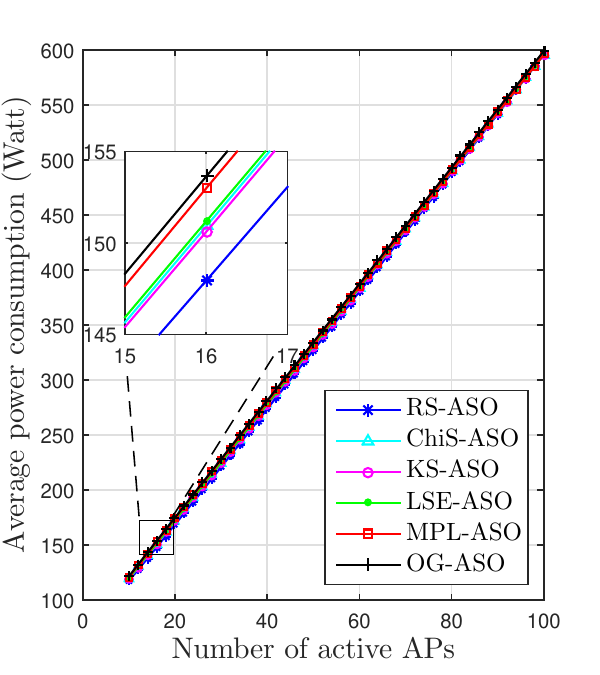}
        \label{fig:Fig_UL_ASO_strategy_PC}
    \end{subfigure}

    \caption{\small Impact of the \gls{ASO} strategy on the \gls{UL} average energy efficiency, spectral efficiency and power consumption as a function of the number of active \glspl{AP}.}
    \label{fig:var_ASO_strategy_UL}
\end{figure*}

\subsection{Impact of the \gls{ASO} strategy}

We start by assessing in this subsection the performance achieved by each of the proposed and considered \gls{ASO} strategies. Towards this end, Fig. \ref{fig:Weighted_ASO_strategy_EE} shows the impact of the \gls{ASO} strategy by depicting the overall average weighted energy efficiency as a function of the number of active \glspl{AP} when a \gls{DL}/\gls{UL} weighting coefficient $\mu=$0.5 (i.e., \gls{DL} and \gls{UL} are given the same importance) has been enforced. It has been assumed that the total number of \glspl{AP} in the system is $M=$100 and each of them is equipped with an $8 \times 1$ \gls{ULA} of vertical half-wave dipoles and $L=$4 RF chains. The results in this figure have been obtained when simultaneously serving $K=$16 \glspl{MS}. As anticipated, the energy efficiency achieved by the \gls{RS-ASO} and \gls{OG-ASO} schemes act, respectively, as lower- and upper-bounds on the performance attained by any of the other considered \gls{ASO} strategies. Note that the proposed \gls{ASO} schemes can be classified in three groups as a function of the system state information they manage. In particular, the \gls{RS-ASO} scheme would be the only member in the first group, comprising those \gls{ASO} strategies that are completely unaware of the network state and thus make blind \gls{AP} switch-on/off decisions. The second group, comprising the goodness-of-fit techniques-based \gls{ASO} schemes (i.e., the \gls{ChiS-ASO}, the \gls{KS-ASO} and the \gls{LSE-ASO}), assume that the spatial distribution of \glspl{MS} on the service coverage area is known, and make only use of \textit{very large-scale} system-state information in the form of the geographical location of the \glspl{AP}. It is remarkable the rather significant performance improvement provided by these methods over the pure \gls{RS-ASO} given their reliance of such a \textit{coarse} network-state information. Note how the achievable energy efficiency increases the more \glspl{AP} are switched-off up to a certain point where an optimum is reached which, for this particular number of \glspl{MS}, is located around $M_A=$24, 20 and 18 active \glspl{AP} for \gls{ChiS-ASO}, \gls{KS-ASO} and \gls{LSE-ASO} strategies, respectively, which shows a significant deviation from the optimum reached when relying on \gls{RS-ASO}, located around $M_A=$34 \glspl{AP}.

The \gls{MPL-ASO} and \gls{OG-ASO} strategies make up the third group, characterized by a certain degree of knowledge of the \textit{short-term} network-state information in the form of the large-scale propagation losses between \glspl{AP} and \glspl{MS}. Note that these losses are tightly connected to the instantaneous positions of the users over the coverage area. The \gls{MPL-ASO} strategy dynamically adapts to \textit{short-term} variations of the spatial distribution of \gls{MS} and, as shown in Fig. \ref{fig:Weighted_ASO_strategy_EE}, greatly outperforms the \gls{ASO} strategies in the first and second groups. The energy efficiency provided by this strategy increases when switching-off some of the \glspl{AP} until only $M_A=$14 \glspl{AP} are left active. Finally, the \gls{OG-ASO} scheme assumes the complete knowledge of all long-term network-state information necessary to calculate the achievable energy-efficiency, including, among others, the channel spatial correlation matrices, the power control matrices or the power consumption metrics and it is seen to outperform the rest of techniques. However, and rather strikingly, the energy-efficiency performance gap between this \textit{idealistic} approach and the much simpler \gls{MPL-ASO} scheme is modest and, moreover, the optimum number of \glspl{AP} to be left active virtually coincides ($M_A=$14 for \gls{MPL-ASO} \textit{vs} $M_A=$13 for \gls{OG-ASO}).

\begin{figure*}[t!]
    \centering
    \begin{subfigure}[t]{0.32\textwidth}
        \centering
        \includegraphics[height=6.8cm]{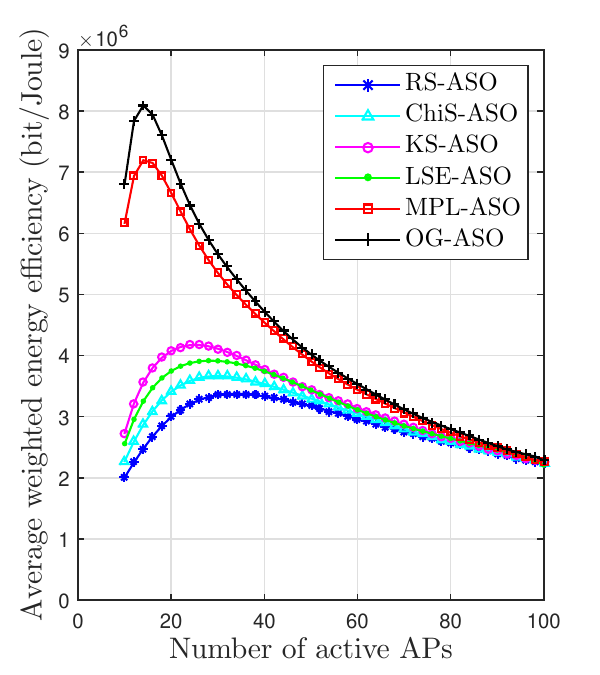}
        \caption{$\mu=$0.5, $\sigma^{\mathcal{S}}$=~0.5297 \\ \text{ }}
        \label{fig:Fig_weigh_ASO_strategy_comm}
    \end{subfigure}
    \begin{subfigure}[t]{0.32\textwidth}
        \centering
        \includegraphics[height=6.8cm]{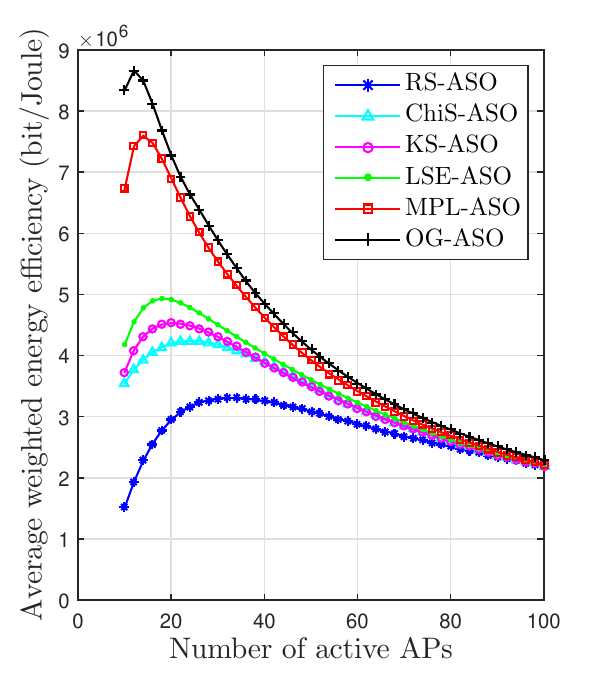}
        \caption{$\mu=$0.5, $\sigma^{\mathcal{S}}$=~2.1188 \\ \text{ }}
        \label{fig:Fig_weigh_ASO_strategy_comm}
    \end{subfigure}
    \begin{subfigure}[t]{0.32\textwidth}
        \centering
        \includegraphics[height=6.8cm]{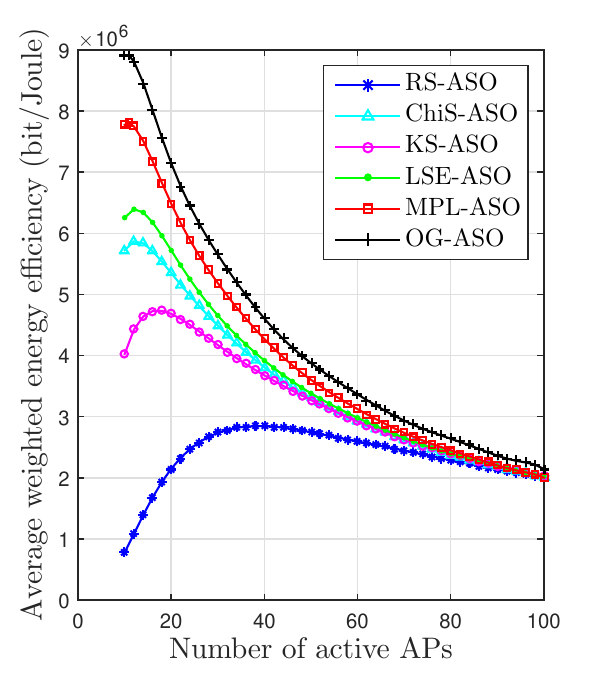}
        \caption{$\mu=$0.5, $\sigma^{\mathcal{S}}$=~8.4752 \\ \text{ }}
        \label{fig:Fig_weigh_ASO_strategy_conc}
    \end{subfigure}
    \caption{\small Average energy efficiency as a function of the number of active \glspl{AP} under scattered (homogeneous), urban and concentrated (hot-spot) spatial distributions of \glspl{MS} (weighted \gls{UL}/\gls{DL} scenario with $\mu=0.5$).}
    \label{fig:conc_ASO_strategy}
\end{figure*}

The energy efficiency, spectral efficiency and power consumption \textit{versus} the number of active \glspl{AP} is presented in Figs. \ref{fig:var_ASO_strategy_DL} and \ref{fig:var_ASO_strategy_UL} for each of the considered schemes and for both \gls{DL} (i.e., $\mu=$0) and \gls{UL} (i.e., $\mu=$1), respectively. Note how now, when considering a pure \gls{DL} setting (Fig. \ref{fig:var_ASO_strategy_DL}), the optimal number of active \glspl{AP} necessary to serve $K=$16 \glspl{MS} is $M_A=$44, 38, 30 and 26 \glspl{AP} for the \gls{RS-ASO}, \gls{ChiS-ASO}, \gls{KS-ASO} and \gls{LSE-ASO} strategies, respectively, and $M_A=$15 and 13 \glspl{AP} for \gls{MPL-ASO} and \gls{OG-ASO} strategies, respectively. In contrast, when focusing on the \gls{UL} case (Fig. \ref{fig:var_ASO_strategy_UL}), the optimal number of active \glspl{AP} necessary to serve $K=$16 \glspl{MS} is $M_A=$23, 14, 13 and 12 \glspl{AP} for the \gls{RS-ASO}, \gls{ChiS-ASO}, \gls{KS-ASO} and \gls{LSE-ASO} strategies, respectively, and $M_A=$10 and 8 \glspl{AP} for \gls{MPL-ASO} and \gls{OG-ASO} strategies, respectively. Jointly considering \gls{DL} and \gls{UL} it is easy to conclude that, irrespective of the \gls{ASO} in use, the \gls{DL} requires of significantly more infrastructure to be active when compared to the \gls{UL}, roughly the double number of active \glspl{AP} if energy efficiency optimality is to be preserved. In fact, it can be generally concluded that for any fixed number of \glspl{AP}, the energy efficiency of the \gls{UL} always exceeds that of the \gls{DL}. This is basically due to two reasons. Firstly, the average power consumption metrics in the \gls{UL} are considerably lower than the corresponding \gls{DL} ones. Secondly, for an optimal number of active \glspl{AP}, the use of full power transmission in the \gls{UL} provides a clear advantage, in terms of spectral efficiency, with respect to the constrained power control transmission implemented in the DL. If a max-min power control strategy was implemented in both \gls{UL} and \gls{UL}, an almost identical spectral efficiency performance would be obtained in both cases (see, for instance, results presented in \cite{Femenias19}), and the energy efficiency advantage shown by the \gls{UL} segment would only be due, in this case, to the lower fixed power consumption.


\subsection{Impact of the distribution of \glspl{MS}}

In order to seize how the concentration of \glspl{MS} influences the decision to choose the most appropriate scheme to maximize the energy efficiency, the effect that parameter $\sigma^{\mathcal{S}}$ has on the network performance is now assessed. In particular, results shown in Fig.~\ref{fig:conc_ASO_strategy} are presented for $\sigma^{\mathcal{S}}$=~2.1188, which corresponds to an \textit{urban common} distribution of \glspl{MS} and that can serve as a baseline against which to compare more \textit{concentrated} distributions ($\sigma^{\mathcal{S}}$=~8.4752) indicating the presence of \textit{hot spots}, and more \textit{scattered} ones ($\sigma^S$=~0.5297), representative of more \textit{homogeneous} suburban environments. Throughout this figure a symmetric UL/DL split is considered ($\mu=$~0.5).

\begin{figure*}[t!]

    \centering
    \begin{subfigure}[t]{0.32\textwidth}
        \centering
        \includegraphics[height=6.8cm]{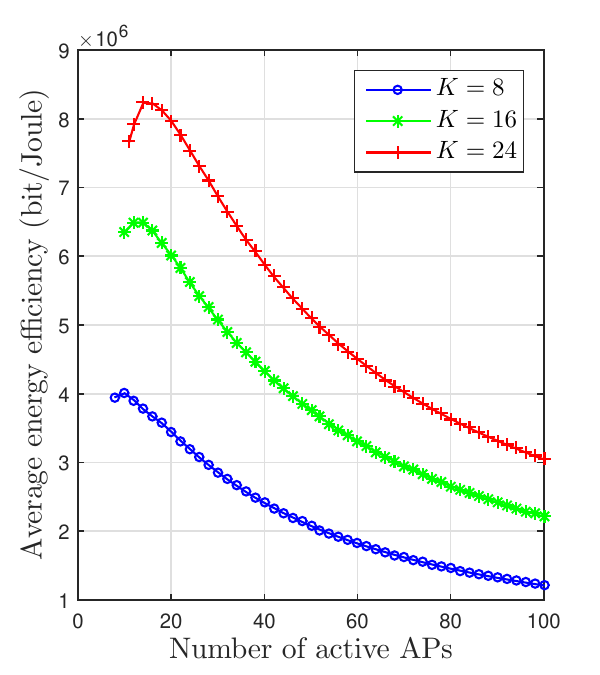}
        \label{fig:Fig_DL_M_K_EE}
    \end{subfigure}
    \begin{subfigure}[t]{0.32\textwidth}
        \centering
        \includegraphics[height=6.8cm]{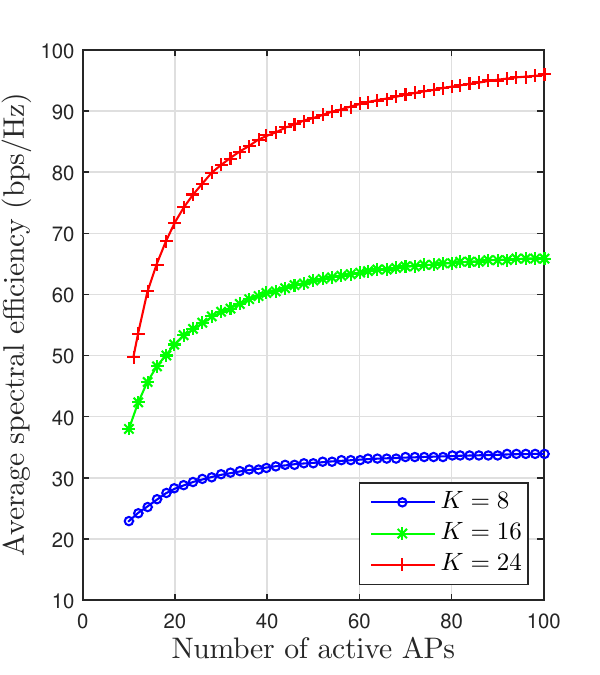}
        \label{fig:Fig_DL_M_K_SE}
    \end{subfigure}
    \begin{subfigure}[t]{0.32\textwidth}
        \centering
        \includegraphics[height=6.8cm]{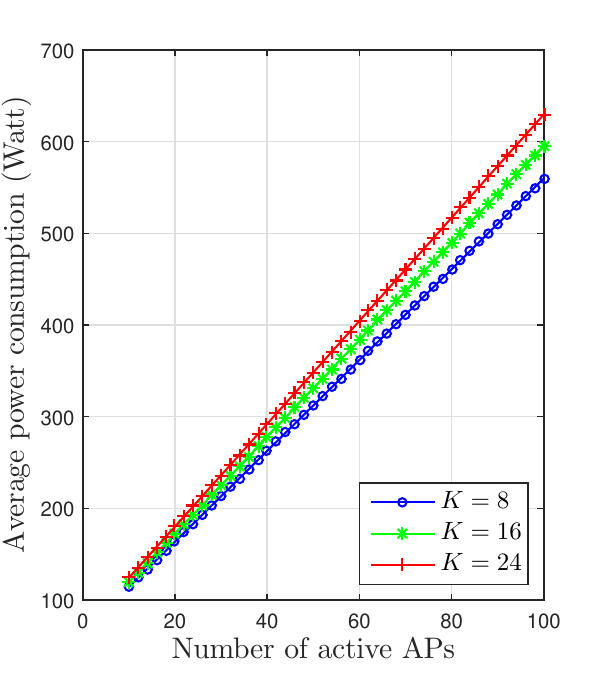}
        \label{fig:Fig_DL_M_K_PC}
    \end{subfigure}

    \caption{\small Impact of the number of \glspl{MS} on the \gls{UL} average energy efficiency, spectral efficiency and power consumption as a function of the number of active \glspl{AP} under the \gls{LSE-ASO} strategy.}
    \label{fig:var_M_K_UL}
\end{figure*}

The first and foremost effect worthwhile pointing out is that \gls{ASO} strategies based on \gls{GoF} offer an energy efficiency performance that greatly exceeds the one attained under the random approach. In fact, note how, as expected, although the performance of the pure \gls{RS-ASO} algorithm is quite acceptable under very homogeneous spatial traffic distributions, it exhibits a clear degradation as the \gls{MS} distribution becomes more heterogeneous (i.e., with increasing $\sigma^{\mathcal{S}}$). Among the \gls{GoF}-based schemes, the \textit{homogeneous} spatial distribution is best discriminated by the \gls{KS-ASO} strategy and the \gls{ChiS-ASO} scheme just offers a slight improvement with respect to the pure random approach. The \gls{LSE-ASO} algorithm is also quite powerful under this circumstances and clearly outperforms the \gls{ChiS-ASO} scheme. As the concentration of \glspl{MS} increases, the behaviours of the \gls{ChiS-ASO} and \gls{KS-ASO} strategies invert. In fact, the sensitivity of the \gls{KS-ASO} scheme to changes in the spatial distribution of \glspl{MS} proves to be very poor. Remarkably, however, although the discrimination power of the \gls{ChiS-ASO} algorithm improves when dealing with \textit{concentrated} spatial distributions of \glspl{MS}, this \gls{ASO} strategy is clearly outperformed by the \gls{LSE-ASO} approach. Summarizing, even though the \gls{KS-ASO} scheme would be the \gls{GoF}-based strategy of choice in scenarios showing a high degree of homogeneity in the spatial distribution of \glspl{MS}, the discrimination power of the \gls{LSE-ASO} approach is the one showing less dependence on the traffic spatial distribution, thus making it the most versatile of them. These results are very consistent with those presented by Aslan and Zech in \cite{Aslan05}.

The \gls{MPL-ASO} and \gls{OG-ASO} techniques, owing to their reliance on far more detailed network information, show a considerable improvement in energy-efficiency performance over the rest of approaches. Moreover, similar to the \gls{GoF}-based techniques, and given their inherent capability to more adequately respond to the spatial distribution of \glspl{MS}, they also reflect a very significant improvement with increasing $\sigma^{\mathcal{S}}$.

Building on the remarks just made, and based on the degree of use of network-state information, performance and complexity, it can be concluded that the \gls{KS-ASO} and \gls{LSE-ASO}, among all the proposed \gls{ASO} strategies, represent the most adequate schemes to be implemented in a cell-free \gls{mmWave} massive \gls{MIMO} system. In particular, \gls{KS-ASO} has shown to be very effective in exploiting mild deviations from \gls{MS} distribution homogeneity ($\sigma^{\mathcal{S}}$~=0.5297), whereas \gls{LSE-ASO} provides the best discrimination power when dealing with spatial distributions of \glspl{MS} showing a marked heterogeneity. Owing to its very good performance across a variety of $\sigma^{\mathcal{S}}$ values, indicative of its robustness, results presented over the next subsections will focus on the use of the \gls{LSE-ASO} scheme. Furthermore, without loss of essential generality, only the optimization of the \gls{UL} segment (i.e., $\mu=$ 1) will be considered (conclusions drawn by using other \gls{ASO} strategies and any other weighting coefficient $\mu$ would be qualitatively equivalent).

\subsection{Impact of the number of \glspl{MS} in the network}

Figure \ref{fig:var_M_K_UL} studies the impact the number of \glspl{MS} has on the \gls{UL} average energy efficiency, spectral efficiency and power consumption as a function of the number of \glspl{AP}. As it can be observed, increasing the number of \glspl{MS} in the network results in an increase in both the average spectral efficiency and the average power consumption. However, the dissimilar increments induced in these two metrics result in quite unalike effects on the average energy efficiency of the network. In particular, fixing the number of active \glspl{AP}, when increasing the network load (i.e., more \glspl{MS}), results in average spectral efficiency increments that more than compensate for the increase in average power consumption, hence raising the average energy efficiency. On the contrary, fixing the number of \glspl{MS}, increments in the number of active \glspl{AP} translate into small improvements in spectral efficiency and a considerable raise of the consumed power, thus resulting in a substantial deterioration of the average energy efficiency of the network. Therefore, when aiming at a high energy efficiency of the a cell-free \gls{mmWave} massive \gls{MIMO} network it is of prime concern that the number of active \glspl{AP} is appropriately adapted to the number of \glspl{MS} to be serviced. This insight is in fact reinforced by noting that the optimal number of active \glspl{AP} increases with the number of \glspl{MS} in the network.

\begin{figure*}[t!]

    \centering
    \begin{subfigure}[t]{0.32\textwidth}
        \centering
        \includegraphics[height=6.8cm]{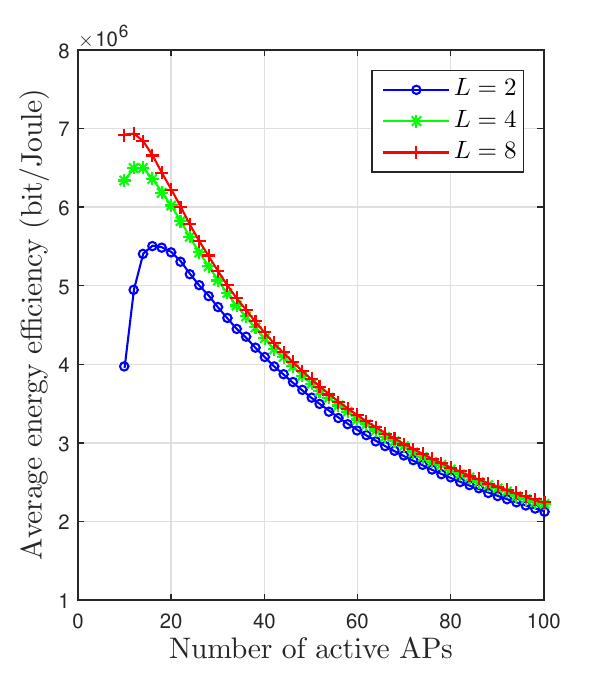}
        \label{fig:Fig_DL_antenna_configuration_ULA_EE}
    \end{subfigure}
    \begin{subfigure}[t]{0.32\textwidth}
        \centering
        \includegraphics[height=6.8cm]{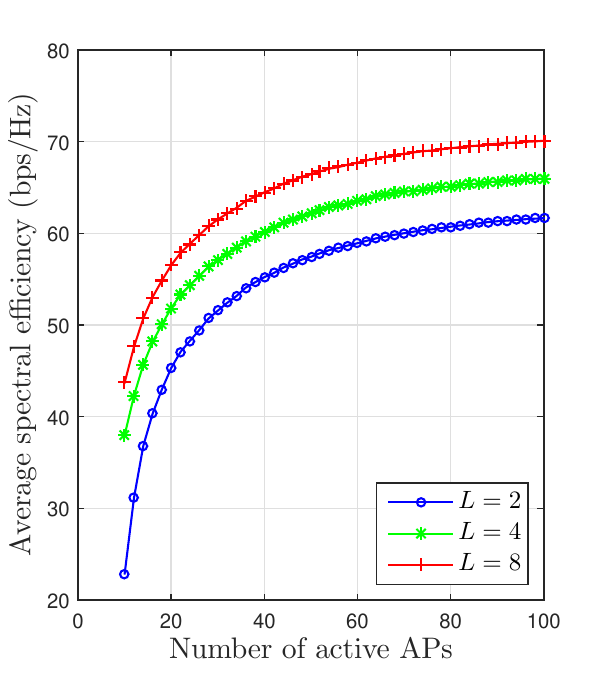}
        \label{fig:Fig_DL_antenna_configuration_ULA_SE}
    \end{subfigure}
    \begin{subfigure}[t]{0.32\textwidth}
        \centering
        \includegraphics[height=6.8cm]{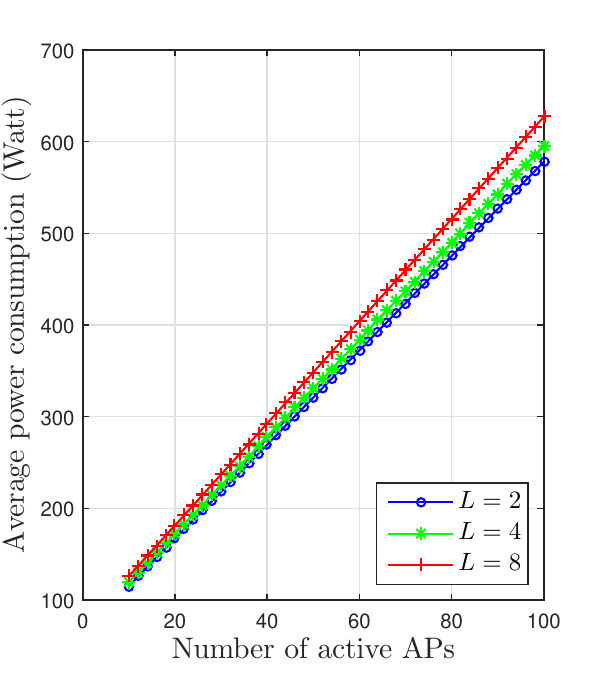}
        \label{fig:Fig_DL_antenna_configuration_ULA_PC}
    \end{subfigure}

    \caption{\small Impact of the \gls{RF} infrastructure used at the \glspl{AP} on the \gls{UL} average energy efficiency, spectral efficiency and power consumption as a function of the number of active \glspl{AP} under the \gls{LSE-ASO} strategy.}
    \label{fig:RF_infrastructure_configuration_UL}
\end{figure*}

\subsection{Impact of the \gls{RF} infrastructure used at the \glspl{AP}}

To understand how the \gls{RF} infrastructure used at the \glspl{AP} influences the performance of the system, Fig. \ref{fig:RF_infrastructure_configuration_UL} represents the energy efficiency, spectral efficiency and power consumption \textit{versus} the number of active \glspl{AP} assuming that each of them is equipped with an $8 \times 1$ linear uniform array and use an analog precoder with $L=$2, 4 and 8 \gls{RF} chains. Note that the latter case indeed corresponds to a system with fully digital processing capability. Results shown next have been obtained assuming the use of a \gls{LSE-ASO} strategy, and the availability of $M=$ 100 \glspl{AP} to serve $K=$16 \glspl{MS}. Naturally, both the average spectral efficiency and the power consumption increase with the number of available \gls{RF} chains. Since the number of active \gls{RF} chains at each of the \glspl{AP} in the network is equal to $L_A = \min\{K,L\}$, increasing the number of available \gls{RF} chains is always beneficial for scenarios where $K \geq L$. Furthermore, note how, as Fig. \ref{fig:var_M_K_UL} reveals, the optimum number of active \glspl{AP} to maximize energy efficiency decreases with the number of \gls{RF} chains available at each \gls{AP}. In particular, for the scenario under consideration, with $L=$2 \gls{RF} chains the optimal number of active \glspl{AP} is equal to $M_A^*=$16, whereas using $L=$8 \gls{RF} chains at each of the \glspl{AP}, the corresponding optimal number of active \glspl{AP} is equal to $M_A^*=$12.

\section{Conclusion}
\label{sec:Conclusion}

This paper has presented a comprehensive analytical framework for the evaluation of the energy efficiency of \gls{AP} sleep-mode techniques for cell-free \gls{mmWave} massive \gls{MIMO} networks with non-uniform spatial traffic density. Based on this framework, different \gls{ASO} strategies have been proposed whose goal is to dynamically turn on/off some of the \glspl{AP} in accordance with metrics related to the spatial distribution of \glspl{MS} in the network with the objective of maximizing the energy efficiency. Towards this end, a realistic model to describe a non-uniform distribution of \glspl{MS} has been included in the analysis that serves to capture the spatial traffic heterogeneity.

Aside from revisiting known \gls{ASO} algorithms in the new context, three novel schemes based on goodness-of-fit techniques (i.e., \gls{ChiS-ASO}, \gls{KS-ASO} and \gls{LSE-ASO}) have been introduced whose rationale is to try to match the spatial distribution of active \glspl{AP} to the one of the \glspl{MS}. Additionally, one more technique, termed \gls{MPL-ASO}, that relies on the \gls{AP}-to-\gls{MS} propagation losses has also been postulated. Remarkably, this new family of \gls{GoF}-based \gls{ASO} strategies (in particular \gls{KS-ASO} and \gls{LSE-ASO}) has been shown to perform considerably better than a pure random procedure while relying only on very large scale information in the form of estimates of the spatial distribution of \glspl{MS} and spatial location of the \glspl{AP}. In particular, \gls{KS-ASO} has shown to be very effective in exploiting mild deviations from \gls{MS} distribution homogeneity, whereas \gls{LSE-ASO} provides the best discrimination power when dealing with spatial distributions of \glspl{MS} showing a marked heterogeneity and, furthermore, its performance is very robust against changes in the spatial distribution of traffic. In turn, the \gls{MPL-ASO} algorithm, while requiring some large-scale information (i.e., \gls{AP}-\gls{MS} large-scale fadings), can neglect extra information the \textit{optimum} \gls{OG-ASO} technique requires (i.e., spatial correlation matrices, power control matrices, power consumption metrics) with only a small performance penalty while being considerably less complex. Results have shown that increasing the network load (more \glspl{MS}) implies activating more \glspl{AP} to attain the optimum point of operation in terms of energy efficiency. The \gls{RF} infrastructure at the \glspl{AP} is also seen to play a key role. In particular, the average energy efficiency increases as the number of active \gls{RF} chains used at the \glspl{AP} grows but, interestingly, this in turns allows optimum operation with a smaller number of active \glspl{AP}.

Future work will concentrate on the use of more sophisticated power control strategies and hybrid analog-digital precoding stages, the study of more reactive \gls{ASO} algorithms as well as its implementation issues, and also, the impact the use of finite-capacity fronthaul links between the \glspl{AP} and the \gls{CPU} may have on the current analytical framework.

\bibliographystyle{IEEEtran}
\bibliography{Cell_free_biblio}

\end{document}

%% file: acronyms.tex
\newacronym{3GPP}{3GPP}{third generation partnership project}
\newacronym{5G}{5G}{fifth generation}
\newacronym{6G}{6G}{sixth generation}
\newacronym{AP}{AP}{access point}
\newacronym{ASO}{ASO}{access point switching}
\newacronym[longplural={angles of arrival}]{AoA}{AoA}{angle of arrival}
\newacronym[longplural={angles of departure}]{AoD}{AoD}{angle of departure}
\newacronym{B5G}{B5G}{beyond 5G}
\newacronym{BCD}{BCD}{block coordinate descend}
\newacronym{BRPA}{BRPA}{balanced random pilot assignment}
\newacronym{BS}{BS}{base station}
\newacronym{CAP}{CAP}{compress-after-precoding}
\newacronym{CB}{CB}{conjugate beamforming}
\newacronym{CBP}{CBP}{compress-before-precoding}
\newacronym{CCDF}{CCDF}{complementary cumulative distribution function}
\newacronym{CDF}{CDF}{cumulative distribution function}
\newacronym{CF}{CF}{cell-free}
\newacronym{CoMP}{CoMP}{coordinated multipoint}
\newacronym{CPA}{CPA}{cluster-based pilot assignment}
\newacronym{CPU}{CPU}{central processing unit}
\newacronym{C-RAN}{C-RAN}{cloud radio access network}
\newacronym{CSI}{CSI}{channel state information}
\newacronym{ChiS}{ChiS}{Chi-square test}
\newacronym{ChiS-ASO}{ChiS-ASO}{ChiS-based \gls{ASO}}
\newacronym{DCPA}{DCPA}{dissimilarity cluster-based pilot assignment}
\newacronym{DL}{DL}{downlink}
\newacronym{GoF}{GoF}{goodness-of-fit}
\newacronym{GOPA}{GOPA}{globally optimal power allocation}
\newacronym{KS}{KS}{Kolmogorov-Smirnov test}
\newacronym{KS-ASO}{KS-ASO}{KS-based \gls{ASO}}
\newacronym{LLH}{LLH}{Log-likelihood test}
\newacronym{LLH-ASO}{LLH-ASO}{LLH-based \gls{ASO}}
\newacronym{LOS}{LOS}{line-of-sight}
\newacronym{LS}{LS}{least-squares}
\newacronym{LSE}{LSE}{logarithmic statistical energy test}
\newacronym{LSE-ASO}{LSE-ASO}{logarithmic statistical energy \gls{ASO}}
\newacronym{MD-ASO}{MD-ASO}{mixture discrepancy-based greedy \gls{ASO}}
\newacronym{MIMO}{MIMO}{multiple-input multiple-output}
\newacronym{M-MIMO}{M-MIMO}{massive MIMO}
\newacronym{MMSE}{MMSE}{minimum mean square error}
\newacronym{mmWave}{mmWave}{millimeter wave}
\newacronym{MRC}{MRC}{maximal ratio combining}
\newacronym{MS}{MS}{mobile station}
\newacronym{MSE}{MSE}{mean square error}
\newacronym{MPL-ASO}{MPL-ASO}{minimum propagation losses-aware \gls{ASO}}
\newacronym{MU-MIMO}{MU-MIMO}{multiuser-MIMO}
\newacronym{NCB}{NCB}{normalized conjugate beamforming}
\newacronym{NMRC}{NMRC}{normalized maximal ratio combiner}
\newacronym{NN-ASO}{NN-ASO}{nearest neighbour-based \gls{ASO}}
\newacronym{NOPA}{NOPA}{normalized optimal power allocation}
\newacronym{NLOS}{NLOS}{non-line-of-sight}
\newacronym{OG-ASO}{OG-ASO}{optimal energy efficiency-based greedy \gls{ASO}}
\newacronym{pdf}{pdf}{probability density function}
\newacronym{QoS}{QoS}{quality of service}
\newacronym{rms}{rms}{root mean square}
\newacronym{RHS}{RHS}{right hand side}
\newacronym{RPA}{RPA}{random pilot assignment}
\newacronym{RS-ASO}{RS-ASO}{random selection \gls{ASO}}
\newacronym{RRM}{RRM}{radio resource management}
\newacronym{RF}{RF}{radio frequency}
\newacronym{SINR}{SINR}{signal-to-interference-plus-noise ratio}
\newacronym{SNR}{SNR}{signal-to-noise ratio}
\newacronym{SOPA}{SOPA}{sequential optimal power allocation}
\newacronym{SR-ASO}{SR-ASO}{spatial regularity-based greedy \gls{ASO}}
\newacronym{TDD}{TDD}{time division duplexing}
\newacronym{UDN}{UDN}{ultra dense network}
\newacronym{ULA}{ULA}{uniform linear array}
\newacronym{UPA}{UPA}{uniform planar array}
\newacronym{UL}{UL}{uplink}
\newacronym{ZF}{ZF}{zero-forcing}